\documentclass[11pt]{article}

\usepackage{url}
\usepackage{verbatim,amsmath, amssymb}
\usepackage{empheq}
\usepackage{mathrsfs}
\usepackage{graphics,amsopn,amstext,amsfonts,color}
\usepackage{bm}
\usepackage{setspace}
\usepackage{subfigure}
\usepackage{graphicx}
\usepackage{epstopdf}
\usepackage[numbers,sort&compress,comma]{natbib}
\usepackage{caption}
\usepackage{multirow}
\usepackage{booktabs}
\usepackage{paralist}
\usepackage{colortbl}
\usepackage{appendix}
\usepackage{color}
\usepackage{xcolor}
\usepackage{wrapfig}
\usepackage{graphics,latexsym,amsfonts,verbatim,amsmath}
\usepackage{algorithm}
\usepackage{algpseudocode}
\usepackage{todonotes}

\def\E{{\mathbb E}}

\def\Pr{{\mathbb{P}}}

\def\Re{\mathbb{R}}

\def\I{\mathbb{I}}

\def\hat{\widehat}

\def \P{\mathcal{P}}

\def \Ze{{\mathbb{Z}}}

\def\P{{\mathcal P}}

\def\C{{\mathcal C}}

\def\Re{{\mathbb R}}

\newcommand{\exclude}[1]{}

\def\CVaR{{\mathbf{CVaR}}}
\def\heur{{\rm{heur}}}

\DeclareMathOperator{\conv}{conv}

\DeclareMathOperator{\diag}{diag}

\newtheorem{Theorem}{Theorem}

\newtheorem{Lemma}{Lemma}

\newtheorem{Proposition}{Proposition}

\newtheorem{Example}{Example}

\usepackage{cleveref}

\crefname{Lemma}{Lemma}{Lemma}
\crefname{Proposition}{Proposition}{Proposition}
\crefname{Theorem}{Theorem}{Theorem}
\crefname{Example}{Example}{Example}
\crefname{Corollary}{Corollary}{Corollary}

\newcommand*{\qed}{\hfill\ensuremath{\square}}

\allowdisplaybreaks

\usepackage{fullpage}
\usepackage{setspace}
\usepackage{authblk}
\newcommand{\proof}{\noindent{\em Proof. }}

\title{Scalable Algorithms for the Sparse Ridge Regression}
\author[1]{Weijun Xie\thanks{Email: wxie@vt.edu.}}
\author[2]{Xinwei Deng\thanks{Email: xdeng@vt.edu.}}
\affil[1]{Department of Industrial and Systems Engineering, Virginia Tech, Blacksburg, VA 24061}
\affil[2]{Department of Statistics, Virginia Tech, Blacksburg, VA 24061}

\begin{document}

\maketitle

\begin{abstract}
Sparse regression and variable selection for large-scale data have been rapidly developed in the past decades.
This work focuses on sparse ridge regression, which {enforces the sparsity by use of the $L_{0}$ norm}.
{
We first prove that the continuous relaxation of the mixed integer second order conic (MISOC) reformulation {using perspective formulation} is equivalent to that of the convex integer formulation proposed in recent work. We also show that the convex hull of the constraint system of MISOC formulation is equal to its continuous relaxation.
Based upon these two formulations {(i.e., the  MISOC formulation and convex integer formulation)}, we analyze two scalable algorithms, the greedy and randomized algorithms, for sparse ridge regression with desirable theoretical properties.}
The proposed algorithms are proved to yield near-optimal solutions under mild conditions.
{We further} propose to integrate the greedy algorithm with the randomized algorithm, which can greedily search the features from the nonzero subset identified by the continuous relaxation of the MISOC formulation.
The merits of the proposed methods are {illustrated through numerical examples} in comparison with several existing ones.
\end{abstract}


\noindent\textit {Approximation Algorithm, Chance Constraint, Conic Program, Mixed Integer, Ridge Regression}

\section{Introduction}

{This paper considers the following optimization problem}:
\begin{align}
v^*=\min_{\bm\beta}\left\{\frac{1}{n}\|\bm{y}-\bm{X}\bm\beta\|_2^2+\lambda \|\bm\beta\|_2^2: \|\bm\beta\|_0\leq k\right\}. \tag{F0}\label{eq_model}
\end{align}
{We refer such an optimization problem as the \textit{sparse ridge regression}, which is also studied by \cite{bertsimas2017sparse,hazimeh2018fast,mazumder2017subset,pilanci2015sparse}. In \eqref{eq_model}, }$\bm y\in \Re^{n}$ denotes the response vector, $\bm X=[\bm x_1,\cdots,\bm x_{p}]\in \Re^{n\times p}$ represents the model matrix,
$\bm\beta \in \Re^{p}$ is the vector of regression coefficients (i.e., estimand),
and $\lambda> 0$ is a positive tuning parameter for the ridge penalty (i.e., {squared} $L_{2}$ penalty). Besides, $\|\bm\beta\|_0$ is the $L_{0}$ norm, which counts the number of nonzero entries of vector $\bm\beta$.
The value of $k$ represents the number of features to be chosen.
In \eqref{eq_model}, {we aim to} find {the} best $k$-sparse estimator, which minimizes the least squares error with a squared $L_2$ penalty.
Without loss of generality, let us assume that $k\leq min(n,p)$. 

{ Note that formulation \eqref{eq_model} is quite general and can be shown to equivalent to the following convex quadratic program with $L_0$ constraint:
\begin{align}
\min_{\bm\beta}\left\{\bm{\beta}^\top \bm{Q}\bm{\beta}-2\bm{a}^\top \bm{\beta}+b: \|\bm\beta\|_0\leq k\right\}, \tag{QP}\label{eq_modelqp}
\end{align}
where $\bm{Q}$ is a symmetric and positive definite matrix. Formulation \eqref{eq_modelqp} is equivalent to \eqref{eq_model} by choosing $\lambda$ to be a positive number that is less than the smallest eigenvalue of $\bm{Q}$ and $\bm{X}=\sqrt{n}(\bm{Q}-\lambda\bm {I})^{1/2},\bm{y}=\sqrt{n}(\bm{Q}-\lambda\bm {I})^{-1/2}\bm{a}, b=\bm{a}^\top (\bm{Q}-\lambda\bm {I})^{-1}\bm{a}$.
}

Sparse ridge regression \eqref{eq_model} can be reformulated as a chance constrained program (CCP) with finite support \citep{ahmed2014nonanticipative,luedtke2014branch}.
That is, we consider $p$ scenarios with equal probability $\frac{1}{p}$, where the $i$th scenario set is $S^i:=\{\bm\beta:\beta_i=0\}$ for $i \in [p]$.
The constraint $\|\bm\beta\|_0 \leq k$ means that at most {$k$ out of the $p$ scenarios} can be violated.
Hence, we can reformulate \eqref{eq_model} as a CCP below：
\begin{align}
v^*=\min_{\bm\beta}\left\{\frac{1}{n}\|\bm y-\bm X\bm\beta\|_2^2+\lambda \|\bm\beta\|_2^2: \frac{1}{p}\sum_{i\in [p]}\I(|\beta_i|\leq 0)\geq 1-\frac{k}{p}\right\}, \tag{F0-CCP}\label{eq_model2}
\end{align}
where $\I(\cdot)$ denotes the indicator function.
In \Cref{sec_existing_method},
we will investigate {the extent of} the recent progress on CCP (e.g., \cite{ahmed2014nonanticipative,luedtke2014branch,nemirovski2006convex}) {which }can be used to solve \eqref{eq_model2}.
It appears that many existing approaches may not work well due to the scalability issue or may result in trivial solutions.
In \Cref{sec_apprximation_alg}, we {conduct and analyze} two scalable algorithms as well as an integration of these two algorithms to solve the sparse ridge regression with theoretical guarantees.

{\it Relevant Literature.}
The ridge regression has been extensively studied in statistics \citep{draper1979ridge,marquardt1975ridge,vinod1978survey}. {It has been shown from existing literature \cite{draper1979ridge,marquardt1975ridge,vinod1978survey} that the additional ridge penalty $\lambda \|\bm\beta\|_2^2$ in \eqref{eq_model} has several desirable advantages including stable solution, estimator variance reduction, and efficient computation. Some recent progress in \cite{blanchet2019robust,gao2017wasserstein} shows that under a certain distributional ambiguity set, the optimal regression coefficients found in \eqref{eq_model} are more robust than those from the conventional sparse regression model, if the data $(\bm{X},\bm{y})$ are insufficient or are subject to some noises.}
However, although {it has} many desirable properties, the ridge estimator is often not sparse. 
Enabling sparsity in regression {has been the focus of} a significant amount of work, including the $L_{1}$ penalty \citep{tibshirani1996regression}, the Bridge estimator using the $L_{q}$ $(q>0)$ penalty \citep{huang2008asymptotic},
the non-convex {SCAD} penalty \citep{xie2009scad}, the minimax concave penalty \citep{zhang2010nearly}, among many others. Several excellent and comprehensive reviews of sparse regression can be found in \cite{buhlmann2011statistics}, \cite{hastie2009unsupervised}, and \cite{fan2011nonconcave}.
In particular, it is worth mentioning that in \cite{zou2005regularization}, the authors proposed a well-known ``elastic net" approach,
which integrates the ridge penalty (i.e., squared $L_2$ penalty) and $L_1$ penalty into the ordinary {least squares} objective to obtain a sparse estimator.
However, similar to the {$L_1$ penalty} method, the elastic net might not consistently find {exactly $k$-sparse estimator}.
On the contrary, instead, we introduce a constraint $\|\bm\beta\|_0\leq k$ in \eqref{eq_model},
which strictly enforces the sparsity on $\bm{\beta}$, and therefore, can obtain the best $k$-sparse estimator.

It has been proven that exact sparse linear regression \eqref{eq_model} {with $\lambda=0$} is NP-hard (cf., \cite{natarajan1995sparse}), so is the sparse ridge regression \eqref{eq_model}.
Various effective approximation algorithms or heuristics {have been} introduced to solve sparse regression \citep{dong2016exact,dong2015regularization,friedman2012fast,kowalski2009sparse,kuo1998variable,miller2002subset}.
For example, in \cite{das2008algorithms}, the authors studied the greedy approach (or forward stepwise selection method) and proved its approximation guarantee when the covariance matrix is nearly identity and has a constant bandwidth. {In \cite{das2011submodular}, the authors} relaxed this assumption and showed that to maximize the $R^2$ statistic for linear
regression, the greedy approach yields a constant approximation ratio {under appropriate conditions}. However, the greedy approach has been found prohibitively expensive when the number of features (i.e., $p$) becomes large \citep{seeger2003fast}.
Recently, 
\cite{hazimeh2018fast} integrated coordinate descent with local combinatorial search, and reported that the proposed method {can numerically} outperform existing ones. However, this method does not provide any provable guarantee on the {global optimality}. Many researchers have also attempted to solve sparse regression by developing exact algorithms (e.g., branch and cut), or using mixed integer program (MIP) solvers. It has been shown that for certain large-sized instances with large signal-to-noise ratios, 
MIP approaches with warm start (a good initial solution) work quite well and can yield very high-quality solutions \citep{bertsimas2016best,bertsimas2017sparse,mazumder2017discrete,mazumder2017subset,miyashiro2015mixed,miyashiro2015subset,atamturk2019rank}. In particular, in \cite{bertsimas2017sparse}, the authors also studied sparse ridge regression and developed a branch and cut algorithm. However, through our numerical study,
these exact approaches can only solve medium-sized instances to near-optimality, and their performances highly rely on the speed of commercial solvers and can vary significantly from one dataset to another.
In this work, our emphasis is to develop fast approximation algorithms with attractive scalability and theoretical performance guarantees.

{\it Our Approaches and Contributions.}
In this work, we will focus on studying sparse ridge regression \eqref{eq_model} and deriving scalable algorithms.
We will first investigate various existing approaches of CCP to solve \eqref{eq_model2}.
One particular approach, which has been used to solve sparse regressions \citep{bertsimas2016best}, is to introduce one binary variable for each indicator function in \eqref{eq_model2} and linearize it with the big-M coefficient.
However, such a method can be very slow in computation, in particular for large-scale datasets.
To overcome the aforementioned challenge,
we develop a \textit{big-M free} mixed integer second order conic (MISOC) reformulation for \eqref{eq_model2}.
We further show that its continuous relaxation is equivalent to that of a mixed integer convex (MIC) formulation in \cite{bertsimas2017sparse,dong2016exact}.
Moreover, these two formulations motivate us to construct a greedy approach (i.e., forward selection) in a much more efficient way than previously proposed in the literature.
The performance guarantee of our greedy approach is also established. A randomized algorithm is studied by investigating the continuous relaxations of the proposed MISOC formulation.
The numerical study shows that the proposed methods work quite well. In particular, the greedy approach outperforms the other methods both in running time and accuracy of variable selection.
The contributions are summarized below:

\begin{enumerate}[(i)]
\item We investigate theoretical properties of three existing approaches of CCP to solve \eqref{eq_model2}, i.e., the big-M method, the conditional-value-at-risk (i.e., $\CVaR$) approach \citep{nemirovski2006convex}, and the heuristic algorithm from \cite{ahmed2014nonanticipative}, and shed some lights on why those methods {may not be amenable} to solve the sparse ridge regression \eqref{eq_model}.

\item We establish a mixed integer second order conic (MISOC) reformulation for \eqref{eq_model2} from perspective formulation \citep{gunluk2012perspective} and prove its continuous relaxation is equivalent to that of a mixed integer convex formulation in the work of \cite{bertsimas2017sparse,dong2016exact}. {We prove that the convex hull of MISOC formulation is equivalent to its continuous relaxation.} We also show that the proposed MISOC formulation can be stronger than the naive big-M formulation.

\item Based on the reformulations, we develop an efficient greedy approach to solve \eqref{eq_model2}, and prove its performance guarantee under a mild condition.
The proposed greedy approach is theoretically sound and computationally efficient.

\item By establishing a relationship between the continuous relaxation value of the MISOC formulation and the optimal value of \eqref{eq_model2} (i.e., $v^*$),
{we analyze a randomized algorithm based on the optimal continuous relaxation solution of the MISOC formulation, and derive its theoretical properties.}
Such a continuous relaxation solution can help reduce the number of potential features and thus can be integrated with the greedy approach.

\item {Our numerical study shows that the proposed methods work quite well, in particular, for the large-scale instances, the proposed greedy  approach can outperforms the others both in running time and accuracy.}
\end{enumerate}

The remainder of the paper is organized as follows.
Section~\ref{sec_existing_method} investigates the applicability of several existing approaches of CCP to solve the sparse ridge regression \eqref{eq_model}.
Section~\ref{eq_formulation} develops two big-M free mixed integer convex program formulations and proves their equivalence.
Section~\ref{sec_apprximation_alg} proposes and analyzes two scalable algorithms and proves their performance guarantees.
{Section~\ref{sec_generalization} introduces the generalized cross validation  to select a proper tuning parameter and a generalization of the proposed formulations to the sparse matrix estimation. }
The numerical experiments of the proposed scalable algorithms are presented in Section~\ref{sec_numerical}. We conclude this work with some discussion in Section~\ref{sec_conclusion}.\\

The following notation is used throughout the paper.
We use bold-letters (e.g., $\bm{x},\bm{A}$) to denote vectors or matrices, and use corresponding non-bold letters to denote their components.
Given a positive integer number $t$, we let $[t]{=\left\{1,\ldots, t\right\}}$ and let $\bm I_{t}$ denote the $t\times t$ identity matrix.
Given a subset $S\subseteq [p]$, we let $\bm\beta_{S}$ denote the subvector of $\bm\beta$ with entries from a subset $S$, and $\bm{X}_{S}$ be a submatrix of $\bm{X}$ with columns from a subset $S$.
For a matrix $\bm{Y}$, we let $\sigma_{\min}(\bm{Y})$ and $\sigma_{\max}(\bm{Y})$ denote its smallest and largest singular values, respectively.
Given a vector $\bm x$, we let $\diag(\bm{x})$ be a diagonal matrix with diagonal entries from $\bm x$.
For a matrix $\bm{W}$, we let $\bm{W}_{\bullet i}$ denotes its $i$th column.
Given a set $T$, we let $\conv(T)$ denote its convex hull.
Given a finite set $S$, we let $|S|$ denote its cardinality.
Given two sets $S,T$, we let $S \setminus T$ denote the set of elements in $S$ but not in $T$, let $S \cup T$ denote the union of $S$ and $T$ and let $S\Delta T$ be their symmetric difference, i.e., $S\Delta T=(S\setminus T)\cup (T\setminus S)$.


\section{{Investigating Existing Solution Approaches on Solving CCP}}\label{sec_existing_method}

In this section, we investigate three \\
commonly-used approaches to solve \eqref{eq_model2}.

\subsection{Big-M Method}

One typical method for a CCP is to formulate it as a mixed integer program (MIP) by introducing a binary variable $z_i$ for each scenario $i\in [p]$, i.e., $\I(\beta_i\neq0) \leq z_i$, and then using big-M method to linearize it, i.e., suppose that $|\beta_i|\leq M_i$ with a large positive number $M_{i}$, then $z_i\geq \I(\beta_i\neq0)$ is equivalent to $|\beta_i| \leq M_iz_i$.
Therefore, \eqref{eq_model2} can be reformulated as the following MIP:
\begin{align}
v^*=\min_{\bm\beta,\bm z}\left\{\frac{1}{n}\|\bm y- \bm X\bm\beta\|_2^2+\lambda \|\bm\beta\|_2^2: \sum_{i\in [p]}z_i \leq k, |\beta_i| \leq M_{i}z_i,\bm z\in \{0,1\}^n\right\}.\tag{F0-big-M} \label{eq_model3} 
\end{align}
{The above formulation \eqref{eq_model3} has been used widely in recent works on sparse regression (see, e.g., \citep{bertsimas2016best,bertsimas2017sparse,mazumder2017discrete,mazumder2017subset,miyashiro2015mixed,miyashiro2015subset}). }The advantage of \eqref{eq_model3} is that it can be directly solved by the off-the-shelf solvers (e.g., CPLEX, Gurobi).
However, one has to choose the vector $\bm{M} = (M_{1}, \ldots, M_{p})^{\top}$ properly.

It is known that \eqref{eq_model3} with big-M coefficients typically has a very weak continuous relaxation value.
Consequently, there has been significant research on improving the big-M coefficients of \eqref{eq_model3}, for example, \cite{ahmed2014nonanticipative,bertsimas2016best,pavlikov:2017,qiu2014covering,song2014chance}. However, the tightening procedures tend to be time-consuming in particular for large-scale datasets.
In Section \ref{eq_formulation}, we will introduce two big-M free MIP formulations, whose continuous relaxation can be proven to be stronger than that of \eqref{eq_model3}.

\subsection{$\CVaR$ Approximation}
Another well-known approximation of CCP is the so-called conditional value at risk ($\CVaR$) approximation (see \cite{nemirovski2006convex} for details),
which is to replace the nonconvex probabilistic constraint by a convex $\CVaR$ constraint.
For the sparse ridge regression in \eqref{eq_model2},
the resulting formulation is
\begin{equation}\label{sp_CVaR}
v^{\CVaR}=\min_{\bm\beta} \left\{ \frac{1}{n}\|\bm y- \bm X\bm\beta\|_2^2+\lambda \|\bm\beta\|_2^2 : \inf_{t}\left[-\frac{k}{p} t+\frac{1}{p}\sum_{i\in [p]}\left(|\beta_i|+t\right)_{+}\right] \leq 0 \right\},
\end{equation}
where $(w)_+=\max(w,0)$.
It is seen that \eqref{sp_CVaR} is a convex optimization problem and provides a feasible solution to \eqref{eq_model2}.
Thus $v^{\CVaR}\geq v^*$.
However, we observe that the only feasible solution to \eqref{sp_CVaR} is $\bm\beta=0$.
\begin{Proposition}\label{example_worst_v_cvar}
The only feasible solution to \eqref{sp_CVaR} is $\bm\beta=0$, i.e., $v^{\CVaR}=\frac{1}{n}\|\bm{y}\|_2^2$.
\end{Proposition}
\proof We first observe that the infimum in \eqref{sp_CVaR} must be achievable. Indeed, $h(t):=-\frac{k}{p} t+\frac{1}{p}\sum_{i\in [p]}(|\beta_i|+t)_{+}$ is continuous and convex in $t$, and $\lim_{t\rightarrow\infty}h(t)=\infty$ and $\lim_{t\rightarrow -\infty}h(t)=\infty$. Therefore, the infimum in \eqref{sp_CVaR} must exist. Hence, in \eqref{sp_CVaR}, we can replace the infimum by the existence operator:
\begin{equation*}
v^{\CVaR}=\min_{\bm\beta} \left\{ \frac{1}{n}\|\bm y- \bm X\bm\beta\|_2^2+\lambda \|\bm\beta\|_2^2 : \exists t, -\frac{k}{p} t+\frac{1}{p}\sum_{i\in [p]}\left(|\beta_i|+t\right)_{+} \leq 0 \right\}.
\end{equation*}
Since $\frac{1}{p}\sum_{i\in [p]}\left(|\beta_i|+t\right)_{+} \geq 0$ and $\frac{k}{p}>0$, therefore, $t\geq 0$, i.e.
\begin{equation*}
v^{\CVaR}=\min_{\bm\beta} \left\{ \frac{1}{n}\|\bm y- \bm X\bm\beta\|_2^2+\lambda \|\bm\beta\|_2^2 : \exists t \geq 0, \frac{p-k}{p} t+\frac{1}{p}\sum_{i\in [p]}|\beta_i|\leq 0 \right\},
\end{equation*}
which implies that $t=0$ and $\beta_i=0$ for each $i\in [p]$.
\qed

Therefore, the $\CVaR$ approach yields a trivial solution for \eqref{eq_model2}.
Hence, it is not a desirable approach, and other alternatives are more preferred.

\subsection{Heuristic Algorithm in \cite{ahmed2014nonanticipative}}
In the recent work of \cite{ahmed2014nonanticipative}, the authors proposed a heuristic algorithm for a CCP with a discrete distribution.
It was reported that such a method could solve most of their numerical instances to near-optimality (i.e., within 4\% optimality gap).
The key idea of the heuristic algorithm in \cite{ahmed2014nonanticipative} is to minimize the sum of infeasibilities for all scenarios when the objective value is upper bounded by $v^U$.
Specifically, they considered the following optimization problem
\begin{align}
\min_{\bm\beta}\left\{\sum_{i\in [p]}|\beta_i| :\frac{1}{n}\|\bm y- \bm X\bm\beta\|_2^2+\lambda \|\bm\beta\|_2^2 \leq v^U\right\}.\label{heuristic}
\end{align}
Let $\bm\beta_U^*$ be an optimal solution to \eqref{heuristic} given an upper bound $v^U$ of $v^*$.
The heuristic algorithm is to decrease the value of $v^U$ if $\|\bm\beta_U^*\|_{0} \le k$, and increase it, otherwise.
This bisection procedure will terminate after a finite number of iterations.
The detailed procedure is described in Algorithm \ref{bisec-heuristic}. Let $v^{\heur} $ denote the output solution from Algorithm \ref{bisec-heuristic}. Then clearly,
\begin{Proposition}
For Algorithm \ref{bisec-heuristic}, the following two properties hold:
\begin{enumerate}[(i)]
\item It terminates with at most $\lfloor\log_2(\frac{\|\bm{y}\|_2^2}{n\hat{\delta}})\rfloor+1$ iterations; and
\item It generates a feasible solution to \eqref{eq_model2}, i.e., $v^*\leq v^{\heur}$.
\end{enumerate}

\end{Proposition}

\proof
\begin{enumerate}[(i)]
\item To prove the first part, \Cref{bisec-heuristic} will terminate if and only if $U-L\leq \hat{\delta}$. And after one iteration, the difference between $U$ and $L$ is halved. Suppose Algorithm \ref{bisec-heuristic} will terminate within at most $T$ steps. Then we must have
\[\frac{\|\bm{y}\|_2^2}{n 2^{T-1}}>\hat{\delta},\]
i.e., $T<1+\log_2\left(\frac{\|\bm{y}\|_2^2}{n\hat{\delta}}\right)$.

\item We start with a feasible solution $\bm\beta=0$ to \eqref{eq_model2}. In \Cref{bisec-heuristic}, we keep track of the feasible solutions from iteration to iteration. Thus, the output of \Cref{bisec-heuristic} is feasible to \eqref{eq_model2}, i.e., $v^*\leq v^{\heur}$.
\end{enumerate}
\qed

%
%

\begin{algorithm}[htbp]
\caption{Heuristic Algorithm in \cite{ahmed2014nonanticipative}}
\label{bisec-heuristic}
\begin{algorithmic}[1]
\State Let $L=0$ and $U =\frac{\|\bm{y}\|_2^2}{n}$ be known lower and upper bounds for \eqref{eq_model2}, let $\hat{\delta} > 0$ be the stopping tolerance parameter.
\While{$U-L > \hat{\delta} $}
\State $q \leftarrow (L+U)/2$.
\State Let $\hat{\bm\beta}$ be an optimal solution of \eqref{heuristic} and set $\hat{z}_i =\I(\hat{\beta}_i=0)$ for all $i
\in [p]$.
\If{$\sum_{i\in [p]}\hat{z}_i \geq p-k$}
\State $U \leftarrow q$.
\Else
\State $L \leftarrow q$.
\EndIf
\EndWhile
\State Output $v^{\heur} \leftarrow U$.
\end{algorithmic}
\end{algorithm}

{It is worth mentioning that for any given upper bound $v^{U}$, the formulation \eqref{heuristic} is similar to the elastic net proposed by \cite{zou2005regularization}, which can be interpreted as a Lagrangian relaxation of \eqref{heuristic}.
The difference between \Cref{bisec-heuristic} and elastic net is that this iterative procedure simultaneously guarantees the sparsity and reduces the regression error while elastic net seeks a trade-off among the regression error, squared $L_2$ penalty, and $L_1$ penalty of $\bm\beta$.
We also note that \Cref{bisec-heuristic} might not be computationally efficient since it requires solving \eqref{heuristic} multiple times but a warm start from the solution of the previous iteration might help speed up the algorithm. Although there have been much development of statistical properties of the elastic net method \cite{zou2005regularization,de2009elastic},
to the best of our knowledge, there is not a known performance guarantee (i.e., approximation ratio) for \Cref{bisec-heuristic}.}

\section{{Investigating Two Big-M Free Reformulations and their Formulation Comparisons}}\label{eq_formulation}

Note that the Big-M formulation in \eqref{eq_model3} is quite compact since it only involves $2p$ variables (i.e., $\bm{\beta},\bm{z}$).
However, it is usually a weak formulation in the sense that the continuous relaxation value of \eqref{eq_model3} can be quite far from the optimal value $v^*$.
In this section, we propose two big-M free reformulations of \eqref{eq_model2}
that arise from two distinct perspectives and prove their equivalence.

\subsection{Mixed Integer Second Order Conic (MISOC) Formulation}
In this subsection, we will {present} a MISOC formulation and its analytical properties.
To begin with, we first make an observation from the perspective formulation in \cite{gunluk2012perspective,ceria1999convex,frangioni2006perspective,dong2015regularization}, {where in \cite{dong2015regularization}, the authors introduced perspective relaxation for sparse regression with $L_0$ penalty term, where they convexified a quadratic term using perspective formulation.}
Let us consider a nonconvex set
\begin{align}
W_i:=\left\{(\beta_i,\mu_i,z_i): \beta_i^2\leq \mu_i, z_i\geq\I(\beta_i\neq 0), z_i\in \{0,1\}\right\},\label{eq_W_i}
\end{align}
for each $i\in [p]$.
The results in \cite{gunluk2012perspective} shows that the convex hull of $W_i$, denoted as $\conv(W_i)$, can be characterized as below.
\begin{Lemma}\label{lem_perp_form}(Lemma 3.1. in \cite{gunluk2012perspective}) For each $i\in [p]$, the convex hull of the set $W_i$ is
\begin{align}
\conv(W_i)=\left\{(\beta_i,\mu_i,z_i): \beta_i^2\leq \mu_iz_i, z_i\in [0,1]\right\}.\label{eq_conv_W_i}
\end{align}
\end{Lemma}

\Cref{lem_perp_form} suggests an extended formulation for \eqref{eq_model2} without big-M coefficients.
To achieve this goal, we first introduce a variable $\mu_i$ to be the upper bound of $\beta_i^2$ for each $i\in [p]$, and a binary variable $z_i\geq \I(\beta_i\neq 0)$. Thus, \eqref{eq_model2} is equal to
\begin{align*}
v^*=\min_{\bm\beta,\bm\mu,\bm{z}}\left\{\frac{1}{n}\|\bm y- \bm X\bm\beta\|_2^2+\lambda \|\bm\mu\|_1: \sum_{i\in [p]}z_i\leq k, (\beta_i,\mu_i,z_i)\in W_i, \forall i\in [p]\right\},
\end{align*}
which can be equivalently reformulated as
\begin{align}
v^*=\min_{\bm\beta,\bm\mu,\bm{z}}\bigg\{\frac{1}{n}\|\bm y- \bm X\bm\beta\|_2^2+\lambda \|\bm\mu\|_1:& (\beta_i,\mu_i,z_i)\in \conv(W_i), z_i\in \{0,1\},\forall i\in [p],\notag\\
& \sum_{i\in [p]}z_i\leq k\bigg\}.\label{eq_model2_persp}
\end{align}
Note that (i) in \eqref{eq_model2_persp}, we replace $W_i$ by $ \conv(W_i)$ and enforce $z_i$ to be binary for each $i\in [p]$; and (ii) from \Cref{lem_perp_form}, $\conv(W_i)$ can be described by \eqref{eq_conv_W_i}.

The above result is summarized in the following theorem.
\begin{Theorem}\label{thm_pers} The formulation \eqref{eq_model2} is equivalent to
\begin{align}
v^*=\min_{\bm\beta,\bm\mu,\bm{z}}\left\{\frac{1}{n}\|\bm y- \bm X\bm\beta\|_2^2+\lambda \|\bm\mu\|_1: \sum_{i\in [p]}z_i\leq k, \beta_i^2\leq \mu_iz_i,z_i\in \{0,1\},\forall i\in [p]\right\}.\tag{F0-MISOC}\label{eq_model2_persp2}
\end{align}
\end{Theorem}
This formulation \eqref{eq_model2_persp2} introduces $p$ more variables $\{\mu_i\}_{i\in [p]}$ than \eqref{eq_model3}, but it does not require any big-M coefficients.

Next, we show that the convex hull of the feasible region of \eqref{eq_model2_persp2} is equal to that of its continuous relaxation. Therefore, it suggests that we might not be able to improve the formulation by simply exploring the constraint system of \eqref{eq_model2_persp2}.
For notational convenience, let $T$ denote the feasible region of \eqref{eq_model2_persp2}, i.e.,
\begin{align}
T=\left\{(\bm\beta,\bm\mu,\bm{z}):\sum_{i\in [p]}z_i\leq k, \beta_i^2\leq \mu_iz_i,z_i\in \{0,1\},\forall i\in [p]\right\}.\label{eq_set_T}
\end{align}
The following result indicates that the continuous relaxation of the set $T$ is equivalent to $\conv(T)$,
\begin{Proposition}\label{cont_rel_T} Let $T$ denote the feasible region of \eqref{eq_model2_persp2}. Then
\[\conv(T)=\left\{(\bm\beta,\bm\mu,\bm{z}):\sum_{i\in [p]}z_i\leq k, \beta_i^2\leq \mu_iz_i,z_i\in [0,1],\forall i\in [p]\right\}.\]
\end{Proposition}
\proof
Let $\hat{T}$ be the continuous relaxation set of $T$, i.e.,
\[\hat{T}=\left\{(\bm\beta,\bm\mu,\bm{z}):\sum_{i\in [p]}z_i\leq k, \beta_i^2\leq \mu_iz_i,z_i\in [0,1],\forall i\in [p]\right\}.\]
We would like to show that $\conv(T)= \hat{T}$. We separate the proof into two steps, i.e., prove  $\conv(T)\subseteq \hat{T}$ and $\hat{T}\subseteq \conv(T)$.
\begin{enumerate}[(i)]
\item It is clear that $\conv(T)\subseteq \hat{T}$.
\item To prove $\hat{T}\subseteq \conv(T)$, we only need to show that for any given point $(\hat{\bm\beta},\hat{\bm\mu},\hat{\bm z}) \in \hat{T}$, we have $(\hat{\bm\beta},\hat{\bm\mu},\hat{\bm z}) \in \conv(T)$. Since $\hat{\bm z}\in \{\bm z:\sum_{i\in [p]}z_i\leq k,\bm{z}\in [0,1]^p\}$, which is an integral polytope, there exists $K$ integral extreme points $\{\bar{\bm z}^t\}_{t\in [K]}\subseteq \Ze_+^p$ such that $\hat{\bm z}=\sum_{t\in[K]}\lambda_t\bar{\bm z}^t$ with $\lambda_t\in (0,1)$ for all $t$ and $\sum_{t\in[K]}\lambda_t=1$. Now we construct $(\bar{\bm\beta}^t,\bar{\bm \mu}^t)$ for each $t\in [K]$ as follows:
\begin{align*}
&\bar{\mu}_{i}^t = \left\{ \begin{array}{ll}
\frac{\hat{\mu}_{i}}{\hat{z}_i} & \text{ if \ } \bar{z}_i^t=1 \\
0 & \text{ otherwise}
\end{array}
\right.,\quad
\bar{\beta}_{i}^t = \left\{ \begin{array}{ll}
\frac{\hat{\beta}_{i}}{\hat{z}_i} & \text{ if \ } \bar{z}_i^t=1 \\
0 & \text{ otherwise}
\end{array}
\right.,\forall i\in[p].
\end{align*}
First of all, we claim that $(\bar{\bm\beta}^t,\bar{\bm\mu}^t,\bar{\bm z}^t)\in T$ for all $t \in [K]$. Indeed, for any $t\in [K]$,
\begin{align*}
&(\bar{\beta}_{i}^t)^2 = \left\{ \begin{array}{ll}
\frac{(\hat{\beta}_{i})^2}{\hat{z}_i^2} & \text{ if \ } \bar{z}_i^t=1 \\
0 & \text{ otherwise}
\end{array}
\right.\leq \bar{\mu}_{i}^t\bar{z}_i^t=\left\{ \begin{array}{ll}
\frac{\hat{\mu}_{i}}{\hat{z}_i} & \text{ if \ } \bar{z}_i^t=1 \\
0 & \text{ otherwise}
\end{array}\right.,\forall i\in[p]\\
&\sum_{i\in [p]}\bar{z}_i^t\leq k\\
& \bar{\bm z}^t\in \{0,1\}^p.
\end{align*}

As $\hat{\bm z}=\sum_{t\in[K]}\lambda_t\bar{\bm z}^t$, thus, for each $i\in [p]$, we have
\begin{align*}
&\sum_{t\in [K]}\lambda_t\bar{\mu}_{i}^t
= \sum_{t\in [K]}\lambda_t\frac{\hat{\mu}_{i}}{\hat{z}_i}\bar{z}_i^t
=\hat{\mu}_{i}\\
&\sum_{t\in [K]}\lambda_t\bar{\beta}_{i}^k
= \sum_{t\in [K]}\lambda_t\frac{\hat{\beta}_{i}}{\hat{z}_i}\bar{z}_i^t
=\hat{\beta}_{i}.
\end{align*}

Thus, $(\hat{\bm\beta},\hat{\bm\mu},\hat{\bm z}) \in \conv(T)$.
\end{enumerate}

\qed

Finally, we remark that if an upper bound $\bm{M}$ of $\bm{\beta}$ is known, then \eqref{eq_model2_persp2} can be further strengthened by adding the constraints ${|\beta_i| \leq M_{i}z_i}$ for each $i\in [p]$. This result is summarized in the following corollary.
\begin{Proposition}\label{thm_pers_cor} The formulation \eqref{eq_model2} is equivalent to
\begin{align}
v^*=\min_{(\bm\beta,\bm\mu,\bm{z})\in T}\left\{\frac{1}{n}\|\bm y- \bm X\bm\beta\|_2^2+\lambda \|\bm\mu\|_1: {|\beta_i| \leq M_{i}z_i},\forall i\in [p]\right\}\tag{F0-MISOC-M}\label{eq_model2_persp3}
\end{align}
where the vector $\bm M = (M_{1}, \ldots, M_{p})^{\top}$ are big-M coefficients and the set $T$ is defined in \eqref{eq_set_T}.
\end{Proposition}
{Please note that the results in \Cref{cont_rel_T} and \Cref{thm_pers_cor} can be generalized to convex quadratic program with side constraints and $L_0$ constraint \cite{bienstock1996computational} such as the portfolio optimization problem.}

\subsection{Mixed Integer Convex (MIC) Formulation}
In this subsection, we will introduce an equivalent MIC formulation to \eqref{eq_model2}. 
The main idea is to separate the optimization in \eqref{eq_model2} into two steps:
(i) we optimize over $\bm\beta$ by fixing its nonzero entries with at most $k$,
and (ii) we select the best subset of nonzero entries with size at most $k$.
After the first step, it turns out that we can arrive at a convex integer program, which is big-M free.
This result has been observed in recent work of \cite{bertsimas2017sparse} and \cite{dong2016exact}.

\begin{Proposition}\label{prop_proj_ref}(\cite{bertsimas2017sparse} and \cite{dong2016exact})
The formulation \eqref{eq_model2} is equivalent to
\begin{align}
v^*&=\min_{\bm z}\left\{f(\bm z):=\lambda \bm{y}^{\top}\left[n\lambda \bm{I}_{n}+\sum_{i\in [p]}z_i\bm{x}_i\bm{x}_i^{\top}\right]^{-1}\bm{y}: \sum_{i\in [p]}z_i \leq k, \bm{z}\in \{0,1\}^p\right\}.\tag{F0-MIC}\label{eq_model6}
\end{align}
\end{Proposition}
Note that in \cite{bertsimas2017sparse}, the authors proposed a branch and cut algorithm to solve \eqref{eq_model6}, which was shown to be effective in solving some large-sized instances.
In the next subsection, we will show that the continuous relaxation of \eqref{eq_model6} is equivalent to that of \eqref{eq_model2_persp2}. Therefore, it can be more appealing to solve \eqref{eq_model2_persp2} directly by MISOC solvers (e.g., CPLEX, Gurobi). Indeed, we numerically compare the branch and cut algorithm with directly solving \eqref{eq_model2_persp2} in \Cref{sec_numerical}.

Finally, we remark that given the set of selected features $S\subseteq [p]$,
its corresponding estimator $\hat{\bm \beta}$ can be computed by the following formula:
\begin{align}
\begin{cases}
\hat{\bm \beta}_S=\left(\bm{X}_S^{\top}\bm{X}_S+n\lambda\bm I_{|S|}\right)^{-1}\bm{X}_S^{\top}\bm y\\
\hat{\beta}_i=0\quad\text{ if }i\in [p]\setminus S
\end{cases}, \label{eq_estimator_S}
\end{align}
where $\hat{\bm \beta}_S$ denotes a sub-vector of $\hat{\bm \beta}$ with entries from subset $S$.

\subsection{Formulation Comparisons}
In this subsection, we will focus on comparing \eqref{eq_model3}, \eqref{eq_model2_persp2}, \eqref{eq_model2_persp3} and \eqref{eq_model6} according to their continuous relaxation bounds.
First, let $v_1,v_2,v_3, v_4$ denote the continuous relaxation of \eqref{eq_model3}, \eqref{eq_model2_persp2}, \eqref{eq_model2_persp3} and \eqref{eq_model6}, respectively, i.e.,
\begin{subequations}
\begin{align}
v_1=&\min_{\bm\beta,\bm z}\left\{\frac{1}{n}\|\bm y- \bm X\bm\beta\|_2^2+\lambda \|\bm\beta\|_2^2: \sum_{i\in [p]}z_i \leq k, |\beta_i| \leq M_{i}z_i,\bm{z}\in[0,1]^p\right\},\label{eq_bound_v1}\\
v_2=&\min_{\bm\beta,\bm\mu,\bm{z}} \left\{\frac{1}{n}\|\bm y- \bm X\bm\beta\|_2^2+\lambda \|\bm\mu\|_1:\beta_i^2 \leq \mu_iz_i,\forall i\in [p],\sum_{i\in [p]}z_i \leq k,\bm{z}\in [0,1]^p\right\},\label{eq_bound_v2}\\
v_3=&\min_{\bm\beta,\bm\mu,\bm{z}} \bigg\{\frac{1}{n}\|\bm y- \bm X\bm\beta\|_2^2+\lambda \|\bm\mu\|_1:\beta_i^2 \leq \mu_iz_i,{|\beta_i| \leq M_{i}z_i},\forall i\in [p],\label{eq_bound_v3}\\
&\qquad\qquad\qquad\qquad\qquad\qquad\quad\sum_{i\in [p]}z_i \leq k,\bm{z}\in [0,1]^p\bigg\},\notag\\
v_4=&\min_{\bm z}\left\{f(\bm z)=\lambda \bm{y}^{\top}\left[n\lambda \bm{I}_{n}+\sum_{i\in [p]}z_i\bm{x}_i\bm{x}_i^{\top}\right]^{-1}\bm{y}: \sum_{i\in [p]}z_i \leq k, \bm{z}\in [0,1]^p\right\}.\label{eq_bound_v4}
\end{align}
\end{subequations}

Next, in the following theorem, we will show a comparison of proposed formulations, i.e., \eqref{eq_model3}, \eqref{eq_model2_persp2}, \eqref{eq_model2_persp3} and \eqref{eq_model6}. In particular, we prove that $v_2 = v_4$, i.e., the continuous relaxation bounds of \eqref{eq_model2_persp2} and \eqref{eq_model6} coincide. In addition, we show that by adding big-M constraints ${|\beta_i| \leq M_{i}z_i}$ for each $i\in [p]$ into \eqref{eq_model2_persp2}, we arrive at a tighter relaxation bound than that of \eqref{eq_model3}, i.e., $v_3\geq v_1$. 
\begin{Theorem}\label{thm_bound_comparison}
Let $v_1,v_2,v_3,v_4$ denote optimal values of \eqref{eq_bound_v1}, \eqref{eq_bound_v2}, \eqref{eq_bound_v3} and \eqref{eq_bound_v4}, respectively. Then
\begin{enumerate}[(i)]
 \item $v_2= v_4 \leq v_3$; and
\item $v_1\leq v_3$.
\end{enumerate}
\end{Theorem}
\proof We separate the proof into three steps.
\begin{enumerate}[(1)]
\item We will prove $v_2= v_4$ first. By Lemma A.1. \citep{sagnol2015computing}, we note that {\eqref{eq_bound_v4}} is equivalent to
\begin{align*}
v_4=&\min_{\bm\gamma_0,\bm\gamma,\bm{z}} \quad \lambda\left( \|\bm\gamma_0\|_2^2+\sum_{i\in [p]} \frac{\gamma_i^2}{z_i}\right),\\
\text{s.t.}\quad& \sqrt{\lambda n}\bm\gamma_0+\sum_{i\in [p]}\bm x_i\gamma_i =\bm y,\\
&\sum_{i\in [p]}z_i\leq k,\\
&\bm{z}\in [0,1]^p, \bm\gamma_0\in \Re^n, \gamma_i\in \Re,\forall i\in [p],
\end{align*}
where by default, we let $\frac{0}{0}=0$. Now let $\beta_i=\gamma_i$ and introduce a new variable $\mu_i$ to denote $\mu_i\geq \frac{\beta_i^2}{z_i}$ for each $i\in [p]$. Then the above formulation is equivalent to
\begin{align*}
v_4=&\min_{\bm\gamma_0,\bm\beta,\bm\mu,\bm{z}}\quad \lambda\left( \|\bm\gamma_0\|_2^2+\|\bm\mu\|_1\right),\\
\text{s.t.}\quad& \sqrt{\lambda n}\bm\gamma_0+\sum_{i\in [p]}\bm x_i\beta_i =y,\\
&\beta_i^2 \leq \mu_iz_i,\forall i\in [p], \\
&\sum_{i\in [p]}z_i\leq k,\\
&\bm{z}\in [0,1]^p, \bm\gamma_0\in \Re^n, \mu_i\in \Re_+,\forall i\in [p].
\end{align*}
Finally, in the above formulation, replace
$$\bm\gamma_0=\frac{1}{\sqrt{\lambda n}}\left(\bm y-\sum_{i\in [p]}\bm x_i\beta_i\right)=\frac{1}{\sqrt{\lambda n}}\left(\bm y-\bm X\bm \beta\right).$$ Then we arrive at \eqref{eq_bound_v2}.

\item Next, we will prove $v_2\leq v_3$. Note that the set of the constraints in \eqref{eq_bound_v2} is a subset of those in \eqref{eq_bound_v3}. Thus, $v_2\leq v_3$.

\item Third, we will prove $v_1\leq v_3$.
{We first note that $v_1$ is equivalent to
\begin{align*}
v_1=&\min_{\bm\beta,\bm\mu,\bm{z}} \bigg\{\frac{1}{n}\|\bm y- \bm X\bm\beta\|_2^2+\lambda \|\bm\mu\|_1:\beta_i^2 \leq \mu_i,{|\beta_i| \leq M_{i}z_i},\forall i\in [p],\\
&\qquad\qquad\qquad\qquad\qquad\qquad\quad\sum_{i\in [p]}z_i \leq k,\bm{z}\in [0,1]^p\bigg\}
\end{align*}
The result $v_1\geq v_3$ follows directly by observing that the constraints $\beta_i^2 \leq \mu_iz_i$ for each $i\in [p]$ imply that $\beta_i^2 \leq \mu_i$ for each $i\in [p]$.
}
\exclude{Let $(\bm\beta^*,\bm\mu^*,\bm z^*)$ be optimal solution of \eqref{eq_bound_v3}. Clearly, we must have $\mu_i^*=\frac{(\beta_i^*)^2}{z_i^*}$ for each $i\in [p]$, otherwise, suppose that there exists $i_0\in [p]$ such that $\mu_{i_0}^*>\frac{(\beta_{i_0}^*)^2}{z_{i_0}^*}$, then the objective value of \eqref{eq_bound_v3} can be strictly less than $v_2$ by letting $\bm\beta=\bm\beta^*$ and $\mu_{i}=\mu_{i}^*$ for each $i\neq i_0$ and $\mu_{i_0} =\frac{(\beta_{i_0}^*)^2}{z_{i_0}^*}$, a contradiction of the optimality of $(\bm\beta^*,\bm\mu^*,\bm z^*)$.

Therefore, we have
\begin{align}
\frac{1}{n}\|\bm y-\bm X\bm\beta^*\|_2^2+\lambda\|\bm\beta^*\|_2^2&\leq \frac{1}{n}\|\bm y-\bm X\bm\beta^*\|_2^2+\lambda \sum_{i\in [p]}\frac{(\beta_i^*)^2 }{z_i^*}\notag\\
&= \frac{1}{n}\|\bm y-\bm X\bm\beta^*\|_2^2+\lambda \|\bm\mu^*\|_1=v_3,\notag
\end{align}
where the inequality is because $z_i^*\in [0,1]$ for each $i\in [p]$. 

Hence, $(\bm\beta^*,\bm{z}^*)$ is feasible to \eqref{eq_bound_v1} with a smaller objective value. Thus, $v_1\leq v_3$.}
 \qed
\end{enumerate}

%
%
%
%
Based on the results established in Theorem \ref{thm_bound_comparison},
we could directly solve the second order conic program \eqref{eq_bound_v2} to obtain the continuous relaxation of MIC \eqref{eq_model6}, which can be solved quite efficiently by existing solvers (e.g., CPLEX, Gurobi).
In addition, adding big-M constraints ${|\beta_i| \leq M_{i}z_i}$ for each $i\in [p]$ into \eqref{eq_bound_v2}, the relaxation bound can be further improved.

Finally, we would like to elaborate that by choosing the vector $\bm{M}$ differently, the continuous relaxation bound $v_2$ of \eqref{eq_model2_persp2} can dominate $v_1$, the continuous relaxation bound of \eqref{eq_model3}, and vice versa.
\begin{Example}\rm Consider the following instance of \eqref{eq_model2} with $n=2,p=2,k=1$ and $\bm y=(1,1)^{\top}, \bm{X}=\bm{I}_2$. Thus, in this case, we have $v^*=\frac{\lambda}{1+2\lambda}+\frac{1}{2}, v_2=\frac{4\lambda}{1+4\lambda}$. There are two different choices about $\bm{M} = (M_{1}, M_{2})^{\top}$:
\begin{enumerate}[(i)]
\item If we choose $\bm{M}$ loosely, i.e., $M_1=M_2=\sqrt{\frac{\|\bm{y}\|_2^2}{n\lambda}}=\sqrt{\frac{1}{\lambda}}$, then
\[v_1=\frac{2\lambda}{1+2\lambda}<v_2< v^*,\]
given that $\lambda>0$.
\item If we choose $\bm{M}$ to be the tightest bound of the optimal solutions of \eqref{eq_model2}, i.e., $M_1=M_2=\frac{1}{1+2\lambda}$, then
\[v_2<v_1=\frac{8\lambda+1}{8\lambda+4}< v^*,\]
given that $\lambda\in (0,1/4)$.
\end{enumerate}
\end{Example}

\section{Two Scalable Algorithms and their Performance Guarantees}\label{sec_apprximation_alg}

In this section,
we will study two scalable algorithms based upon two equivalent formulations \eqref{eq_model2_persp2} and \eqref{eq_model6}, i.e., the greedy approach based on \eqref{eq_model6}, and the randomized algorithm based on \eqref{eq_model2_persp2}.

\subsection{The Greedy Approach based on MIC Formulation}
The greedy approach (i.e., forward selection) has been commonly used as a heuristic to conduct the best subset selection \citep{das2011submodular,smola2001sparse,zhang2011adaptive}.
The idea of the greedy approach is to select the feature that minimizes the marginal decrement of the objective value in \eqref{eq_model6} at each iteration until the number of selected features reaches $k$. Note that given a selected subset $S\subseteq [p]$ and an index $j\notin S$, the marginal objective value difference by adding $j$ to $S$ can be computed explicitly via the Sherman-Morrison formula \citep{sherman1950adjustment} as below:
\begin{align*}
&\lambda \bm{y}^{\top}\left[\bm A_S+\bm x_j\bm x_j^{\top}\right]^{-1}\bm y
-\lambda \bm{y}^{\top}A_S^{-1}\bm y
=-\frac{\lambda\left(\bm{y}^{\top}\bm A_S^{-1}\bm x_j\right)^2}{1+\bm x_j^{\top}\bm A_S^{-1}\bm x_j},\\
& \left [ \bm A_S+\bm x_j\bm x_j^{\top} \right]^{-1} =\bm{A}_S^{-1}-\frac{\bm{A}_S^{-1}\bm x_{j}\bm x_{j}^{\top}\bm{A}_S^{-1}}{1+\bm x_j^{\top}\bm{A}_S^{-1}\bm x_j},
\end{align*}
where $\bm A_S=n\lambda \bm{I}_{n}+\sum_{i\in S}\bm{x}_i\bm{x}_i^{\top}$.\\

This motivates us an efficient implementation of the greedy approach, which is described in~\Cref{greedy_approach}. Note that in \Cref{greedy_approach}, at each iteration, we only need to keep track of $\{\bm A_S^{-1}\bm x_j\}_{j \in [p]}$, $\{\bm x_j\bm A_S^{-1}\bm x_j\}_{j \in [p]}$ and $\{\bm y\bm A_S^{-1}\bm x_j\}_{j \in [p]}$, which has space complexity $O(np)$ and update them from one iteration to another iteration, which costs $O(np)$ operations per iteration. Therefore, the space and time complexity of \Cref{greedy_approach} are $O(np)$ and $O(npk)$, respectively.

\begin{algorithm}[htbp]
\caption{Proposed Greedy Approach for Solving \eqref{eq_model6} }
\label{greedy_approach}
\begin{algorithmic}[1]
\State Initialize $S=\emptyset$ and $\bm{A}_S=n\lambda \bm{I}_{n}$
\For{$i=1,\ldots,k$}
\State Let $j^*\in \arg\min_{j\in [p]\setminus S}\left\{ -\frac{\lambda\left(\bm{y}^{\top}\bm{A}_S^{-1}\bm x_j\right)^2}{1+\bm x_j^{\top}\bm{A}_S^{-1}\bm x_j}\right\}$
\State Let $S=S\cup\{j^*\}$ and $\bm{A}_S=\bm{A}_S+\bm x_{j^*}\bm x_{j^*}^{\top}, \bm{A}_S^{-1}=\bm{A}_S^{-1}-\frac{\bm{A}_S^{-1}\bm x_{j^*}\bm x_{j^*}^{\top}\bm{A}_S^{-1}}{1+\bm x_{j^*}^{\top}\bm{A}_S^{-1}\bm x_{j^*}}$
\EndFor
\State Output $v^{G} \leftarrow \lambda \bm{y}^{\top}\bm{A}_S^{-1}\bm y$.
\end{algorithmic}
\end{algorithm}

{From our empirical study, the greedy approach work quite {well}. Indeed, we will investigate the greedy solution and prove that it can be very close to the true optimal, in particular when $\lambda$ is not too small. To begin with, let us define $\theta_s$ to be the largest singular value of all the matrices $\bm{X}_S\bm{X}_S^{\top}$ with $|S|=s$, i.e.,
\begin{equation}
\theta_s:=\max_{|S|= s}\sigma_{\max}^2(\bm{X}_S)=\max_{|S|=s}\sigma_{\max}(\bm{X}_S\bm{X}_S^{\top}), \label{eq_kappa}
\end{equation}
for each $s\in [p]$. By definition \eqref{eq_kappa}, we have $\theta_1\leq \theta_2\leq \ldots\leq\theta_{p}$, and by default, we let $\theta_0=0$.


Our main results of near-optimality of the greedy approach are stated as below.
That is, if $p\geq k$, then the solution of greedy approach will be quite close to any optimal estimator from \eqref{eq_model2} as $\lambda$ grows.
\begin{Theorem}\label{thm_asym_greedy} Suppose $p\geq k$. Then the output (i.e., $v^G$) of the greedy approach (i.e., Algorithm~\ref{greedy_approach}) is bounded by
\begin{align}
v^*\leq v^G \leq \frac{n\lambda+\theta_k}{n\lambda}\left(1-\frac{n^2\lambda^2\underline{\theta}}{(n\lambda+\theta_1)(n\lambda+\theta_k)^2}\log \left(\frac{p+1}{p+1-k}\right)\right)v^*,
\end{align}
where $\theta$ defined in \eqref{eq_kappa} and
\[\underline{\theta}=\min_{T\subseteq [p], |T|\geq p-k+1}\sigma_{\min}(\bm{X}_T\bm{X}_T^\top).\]
\end{Theorem}
\proof

First of all, suppose that $\bm{z}^*$ is an optimal solution to \eqref{eq_model6}. According to the definition of $\theta_k$, we have $n\lambda \bm I_{n}+\sum_{i\in [p]}z_i^*\bm{x}_i\bm{x}_i^{\top} \leq (n\lambda+\theta_k)\bm I_n$. Thus,
\begin{align}
v^*=\lambda \bm y^\top \left(n\lambda \bm I_{n}+\sum_{i\in [p]}z_i^*\bm{x}_i\bm{x}_i^{\top}\right)\bm y&\geq \frac{\lambda}{n\lambda+\theta_k}\|\bm{y}\|_2^2.\label{eq_opt_bound_error}
\end{align}

On the other hand, according to Step 3 of Algorithm~\ref{greedy_approach}, for any given $S$ such that $|S|=s<k$, and $\bm{A}_S=n\lambda \bm I_{n}+\sum_{i\in S}\bm{x}_i\bm{x}_i^{\top}$ and $j\in [p]\setminus S$, we observe that
\begin{align}
&\lambda \bm{y}^{\top}\left[\bm{A}_S+\bm{x}_j\bm{x}_j^{\top}\right]^{-1}\bm{y}-\lambda \bm{y}^{\top}\bm{A}_S^{-1}\bm y=-\frac{\lambda\left(\bm{y}^{\top}\bm{A}_S^{-1}\bm x_j\right)^2}{1+x_j^{\top}\bm{A}_S^{-1}x_j}.\label{eq_greedy_bound_error}
\end{align}

Thus, using the identity \eqref{eq_greedy_bound_error}, we can prove by induction that the greedy value is upper bounded by
\begin{align}
v^G&\leq \left(1-\frac{n^2\lambda^2\underline{\theta}}{(n\lambda+\theta_1)(n\lambda+\theta_k)^2}\sum_{i\in [k]}\frac{1}{p+1-i}\right)\frac{1}{n}\|\bm{y}\|_2^2. \label{eq_grd_ub}
\end{align}
Indeed, if $k=0$, then \eqref{eq_grd_ub} holds. Suppose that $k=t\geq 0$, \eqref{eq_grd_ub} holds. Now let $k=t+1$ and let $S$ be the selected subset at iteration $t$. By induction, we have
\[\lambda \bm{y}^{\top}\bm{A}_S^{-1}\bm y \leq \left(1-\frac{n^2\lambda^2\underline{\theta}}{(n\lambda+\theta_1)(n\lambda+\theta_k)^2}\sum_{i\in [t]}\frac{1}{p+1-i}\right)\frac{1}{n}\|\bm{y}\|_2^2.\]
And by the greedy selection procedure, we further have
\begin{align*}
v^G&=\lambda \bm{y}^{\top}\bm{A}_S^{-1}\bm y+\min_{j\in [p]\setminus S}\lambda \bm{y}^{\top}\left[\bm{A}_S+\bm{x}_j\bm{x}_j^{\top}\right]^{-1}y-\lambda \bm{y}^{\top}\bm{A}_S^{-1}\bm y\\
&\leq \lambda \bm{y}^{\top}\bm{A}_S^{-1}\bm y+\frac{1}{p-t}\sum_{j\in [p]\setminus S}\left[-\frac{\lambda\left(\bm{y}^{\top}\bm{A}_S^{-1}\bm x_j\right)^2}{1+x_j^{\top}\bm{A}_S^{-1}x_j}\right]\\
&\leq \lambda \bm{y}^{\top}\bm{A}_S^{-1}\bm y-\frac{n\lambda^2}{(p-t)(n\lambda+\theta_1)}\bm y^\top \bm A_S^{-1}(\bm{X}_{[p]\setminus S}\bm{X}_{[p]\setminus S}^\top)\bm A_S^{-1}\bm{y}\\
&\leq \left(1-\frac{n^2\lambda^2\underline{\theta}}{(n\lambda+\theta_1)(n\lambda+\theta_k)^2}\sum_{i\in [t+1]}\frac{1}{p+1-i}\right)\frac{1}{n}\|\bm{y}\|_2^2,
\end{align*}
where the first equality is due to \eqref{eq_greedy_bound_error}, the first inequality is because the minimum is no larger than the average, the second inequality is because $\bm A_S\succeq n\lambda \bm I_n$ and $\|\bm x_j\|_2^2\leq \theta_1$, and the third inequality is due to the induction and the facts that $p\geq k, A_S\preceq (n\lambda+\theta_k) \bm I_n, \underline{\theta}\leq \sigma_{\min}(\bm{X}_{[p]\setminus S}\bm{X}_{[p]\setminus S}^\top)$.

Combining \eqref{eq_opt_bound_error} and \eqref{eq_greedy_bound_error} and using the fact that $\sum_{i\in [k]}\frac{1}{p+1-i}\geq \int_{0}^k \frac{1}{p+1-t}dt=\log \left(\frac{p+1}{p+1-k}\right)$, the conclusion follows.
\qed}

{We make the following remarks about Theorem \ref{thm_asym_greedy}.
\begin{enumerate}[(i)]
\item If $p< n+k$, then according to the definition, $\underline{\theta}=0$.
\item If we normalize $\|\bm{x}_i\|_2^2=n$ for each $i\in [p]$, we must have $\theta_k\leq  kn$, thus $\frac{n\lambda}{n\lambda+\theta_k}\leq \frac{\lambda}{\lambda+k}$. Therefore, we can see that the objective value of greedy approach is closer to the true optimal value if the tuning parameter becomes larger.
\item Besides, our analysis and asymptotic optimality of the greedy approach is new without any assumption on the data and thus is quite different from the existing ones for sparse regression \cite{das2008algorithms,das2011submodular,khanna2017scalable,zhang2011adaptive,candes2007dantzig,candes2008restricted}. {For example, the results in \citep{candes2007dantzig,candes2008restricted} require the well-known restricted isometry property (RIP) states as below: \[(1-\delta_s)\|\bm\beta\|_2^2\leq \|\bm{X}\bm\beta\|_2^2 \leq (1+\delta_s) \|\bm\beta\|_2^2, \forall s\in [p],\bm\beta: \|\bm\beta\|_0=s,\] where $\bm\delta\in (0,1)^p$ is a constant. This is quite a strong assumption and our Theorem \ref{thm_asym_greedy} does not require such an assumption.} On the other hand, if the tuning parameter $\lambda\rightarrow 0_+$, then our performance guarantee can be arbitrarily bad. Therefore, our analysis cannot trivially extend to sparse regression.
\end{enumerate}}
In the next subsection, we will {investigate} a randomized algorithm and prove its approximation guarantee under a weaker condition of $\lambda$.

In addition, we remark that the estimator $\bm{\beta}^G$ of the greedy approach can be computed by \eqref{eq_estimator_S}, where $S$ denotes the set of features selected by the greedy approach.
In the next theorem, we will show that the derived estimator from the greedy approach (i.e., $\bm{\beta}^G$) can be also quite close to an optimal solution $\bm\beta^*$ of \eqref{eq_model2}.
{\begin{Theorem}\label{eq_bound_beta_S}
Let $\bm\beta^*$ be an optimal solution to \eqref{eq_model2} with set of selected features $S^*$ and $\bm{\beta}^G$ be the estimator from the greedy approach with set of selected features $S^G$.
Suppose that $p\geq k$, 
then we have
\begin{align*}
\|\bm{\beta}^G-\bm\beta^*\|_2 \leq
\frac{ \sqrt{ 4 n \theta_{|S^G\setminus S^*|}v^* } }{n \lambda +
\sigma_{\min}( \bm{X}_{S^{U}}^{\top} \bm{X}_{S^{U}})} + \sqrt{\frac{n\nu v^*}{n\lambda+\sigma_{\min}( \bm{X}_{S^{U}}^{\top} \bm{X}_{S^{U}})}},
\end{align*}
where $S^{U} = S^G\cup S^*$, i.e., the union of set $S^{G}$ and set $S^{*}$, and
\[\nu=\frac{n\lambda+\theta_k}{n\lambda}\left(1-\frac{n^2\lambda^2\underline{\theta}}{(n\lambda+\theta_1)(n\lambda+\theta_k)^2}\log \left(\frac{p+1}{p+1-k}\right)\right)-1.\]
\end{Theorem}
\proof
Note that the greedy estimator $\bm{\beta}^G$ can be computed through \eqref{eq_estimator_S} by setting $S$ to be $S^{G}$, the set of selected features by greedy approach.
Moreover, we define $\tilde{\bm X}$ as follows:
\[\begin{cases}
\tilde{\bm X}_{S^G\setminus S^*}=\bm X_{S^G\setminus S^*}\\
\tilde{\bm X}_{\bullet i}=0\quad\text{ if }i\in [p]\setminus (S^G\setminus S^*)
\end{cases}. \]
Then we have,
\begin{align*}
&\frac{1}{n}\|\bm y-\bm X\bm{\beta}^G\|_2^2+\lambda \|\bm{\beta}^G\|_2^2 -\left[\frac{1}{n}\|\bm y-\bm X\bm\beta^*\|_2^2+\lambda \|\bm\beta^*\|_2^2\right]\leq \nu v^* \\
(\Leftrightarrow)& -2\left(\bm\beta^*-\bm{\beta}^G\right)^{\top}\left[-\frac{1}{n}\bm X^{\top}\left(\bm y-\bm X\bm\beta^*\right)+\lambda\bm\beta^*\right] \\
&+\left(\bm\beta^*-\bm{\beta}^G\right)^{\top}\left[\frac{1}{n}\bm{X}^{\top}\bm{X}+\lambda\bm I_{p}\right]\left(\bm\beta^*-\bm{\beta}^G\right) \leq \nu v^*\\
(\Leftrightarrow)& -2\left(\bm\beta^*-\bm{\beta}^G\right)^{\top}\left[-\frac{1}{n}\tilde{\bm X}^{\top}\left(\bm y-\bm X\bm\beta^*\right)\right] \\
&+\left(\bm\beta_{S^{U}}^*-\bm{\beta}_{S^{U}}^G\right)^{\top}\left[\frac{1}{n}\bm{X}_{S^{U}}^{\top}\bm{X}_{S^{U}}+\lambda\bm I_{|S^{U}|}\right]\left(\bm\beta_{S^{U}}^*-\bm{\beta}_{S^{U}}^G\right) \leq \nu v^*\\
(\Rightarrow)& -\frac{2}{n}\|\tilde{\bm X}\|_2\|\bm y-\bm X\bm\beta^*\|_2\|\bm\beta_{S^G\setminus S^*}^*-\bm{\beta}_{S^G\setminus S^*}^G\|_2
\\
&+\left(\lambda+\frac{\sigma_{\min}( \bm{X}_{S^{U}}^{\top} \bm{X}_{S^{U}})}{n}\right)\left\|\bm\beta^*-\bm{\beta}^G\right\|_2^2\leq \nu v^* \\
(\Rightarrow)& -\sqrt{\frac{4\theta_{|S^G\setminus S^*|}v^*}{n}}\|\bm\beta^*-\bm{\beta}^G\|_2
+\left(\lambda+\frac{\sigma_{\min}( \bm{X}_{S^{U}}^{\top} \bm{X}_{S^{U}})}{n}\right)\left\|\bm\beta^*-\bm{\beta}^G\right\|_2^2\leq \nu v^* \\
(\Rightarrow)&\left\| \bm{\beta}^G-\bm\beta^*\right\|_2\\
&\leq
\frac{ \sqrt{ 4 n \theta_{|S^G\setminus S^*|}v^* } }{n \lambda + \sigma_{\min}( \bm{X}_{S^{U}}^{\top} \bm{X}_{S^{U}})} + \sqrt{\frac{n\nu v^*}{n\lambda+\sigma_{\min}( \bm{X}_{S^{U}}^{\top} \bm{X}_{S^{U}})}},
\end{align*}
where the second equivalence is due to the optimality condition of $\beta^*$, i.e., \\$-\frac{1}{n}\bm X_{S^*}^{\top}\left(\bm y-\bm X_{S^*}\bm\beta_{S^*}^*\right)+\lambda\bm\beta_{S^*}^*=0$, and the nonzero entries of $\bm\beta^*-\bm{\beta}^G$ are only from subset $S^U:=S^G\cup S^*$.
The first implication is due to sub-multiplicativity of matrix norm and $\|\bm A\|_2\geq \sigma_{\min}(\bm A)$,
the second implication is because of $\|\tilde{\bm X}\|_2\leq \sqrt{\theta_k}, \|\bm y-\bm X\bm\beta^*\|_2\leq \sqrt{nv^*}$,
and the last implication is because any solution of the following quadratic inequality $a t^2-bt-c\leq 0$ with $a,b,c>0$ is upper bounded by $\frac{b}{a}+\sqrt{\frac{c}{a}}$.
\qed}

Note that in \Cref{eq_bound_beta_S}, the first term of the error bound vanishes when $S^G=S^*$, i.e., when the greedy approach can exactly identify all the features.

\subsection{The Randomized Algorithm based on MISOC Formulation}
In this subsection, we {investigate} a randomized algorithm based on the continuous relaxation solution of \eqref{eq_model2_persp2}, i.e., the optimal solution to \eqref{eq_bound_v2},
which can be efficiently solved via the interior point method or other convex optimization approaches \citep{bental2001Lectures}.

Suppose that $\hat{\bm z}$ is the optimal solution of the continuous relaxation model \eqref{eq_bound_v2}.
For each $i\in [p]$, the column $\bm x_i$ will be picked by probability $\hat{z}_i$. The detailed implementation is illustrated in \Cref{alg_rand_round}. 
\begin{algorithm}
\caption{Proposed Randomized Algorithm}
\label{alg_rand_round}
\begin{algorithmic}[1]
\State Let $\hat{\bm z}$ be the optimal solution to \eqref{eq_bound_v2}
\State Initialize set $S=\emptyset$ and vector $\tilde{\bm z}=0\in \Re^p$
\For{$i=1,\ldots,p$}
\State Sample a standard uniform random variable $U$
\If{$U\leq \hat{z}_i$}
\State Let $S=S\cup\{i\}$ and $\tilde{z}_i=1$
\EndIf
\EndFor
\State Output $S, \tilde{\bm z}$
\end{algorithmic}
\end{algorithm}

Next, we will show that if $\lambda$ is not too small, then with high probability,
the output $S$ of \Cref{alg_rand_round} yields its corresponding objective value close to the optimal value $v^*$.
To begin with, we present the following matrix concentration bound.
\begin{Lemma}\label{lemm_trop}(Theorem 1.4., \cite{tropp2012user})
Consider a finite sequence $\{\bm Y_k\}$ of independent, random, symmetric
matrices with dimension $d$. Assume that each random matrix satisfies
$E[\bm Y_k] = 0$ and $\|\bm Y_k\|_2^2\leq R^2$ almost surely.
Then, for all $t\geq 0$, we have
\begin{align}
\Pr\left\{\left\|\sum_{k}\bm Y_k\right\|_2\geq t\right\}\leq d\exp\left(-\frac{t^2}{2\nu^2+2/3Rt}\right),
\end{align}
where $\nu^2:=\|\sum_{k}\E[\bm Y_k^2]\|_2$.
\end{Lemma}

\Cref{lemm_trop} implies that if $\lambda$ is not too small, then with high probability, $\lambda n\bm{I}_{n}+ \sum_{i\in S}\bm{x}_i\bm{x}_i^{\top}$ has the similar eigenvalues as $\lambda n\bm{I}_{n}+ \sum_{i\in [p]}\hat{z}_i\bm{x}_i\bm{x}_i^{\top}$, where $\hat{\bm z}$ is the optimal solution to \eqref{eq_bound_v2} and $S$ is the output of \Cref{alg_rand_round}.
{\begin{Lemma}\label{lem1}Let $\hat{\bm z}$ be the optimal solution to \eqref{eq_bound_v2} and $S$ be the output of \Cref{alg_rand_round}. Given that $\alpha\in (0,1)$ and \[\lambda\geq \frac{\log(2n/\alpha)\sqrt{\theta_1}}{3n\epsilon}+\frac{\sqrt{2\theta_k\log(2n/\alpha)}}{2n\epsilon},\]
then with probability at least $1-\frac{\alpha}{2}$, we have
\begin{align*}
(1-\epsilon)\bm u^{\top}\bm{\Sigma}_*\bm u\leq \bm u^{\top}\hat{\bm\Sigma} \bm u \leq (1+\epsilon)\bm u^{\top}\bm{\Sigma}_*\bm u, \forall \bm u\in \Re^n,
\end{align*}
where $\bm{\Sigma}_*=\lambda n\bm{I}_{n}+ \sum_{i\in [p]}\hat{z}_i\bm{x}_i\bm{x}_i^{\top}$ and $\hat{\bm\Sigma}=\lambda n\bm{I}_{n}+ \sum_{i\in S}\bm{x}_i\bm{x}_i^{\top}$.
\end{Lemma}
\proof
Let $\hat{\bm z}$ be the optimal solution to \eqref{eq_bound_v2} and let $\{r_i\}_{i\in [p]}$ be independent Bernoulli random variables with $\Pr\{r_i=1\}=\hat{z}_i$ for each $i\in [p]$. Consider the random matrix defined as for each $i\in [p]$,
\[\bm{A}_i=(r_i-\hat{z}_i)\bm{x}_i\bm{x}_i^{\top}\]
and $\E[\bm{A}_i]=0$. On the other hand, by definition, we have $\|\bm{x}_i\|_2^2\leq \theta_1$ for each $i\in [p]$, thus
\[\|\bm{A}_i\|_2=|r_i-\hat{z}_i|\|\bm{x}_i\|_2^2\leq \theta_1:=R^2.\]

Also,
\begin{align*}
&\left\|\sum_{i\in[p]}\E[\bm{A}_i^2]\right\|_2=
\left\|\sum_{i\in[p]}\hat{z}_i\left(1-\hat{z}_i\right)\|\bm{x}_i\|_2^2\bm{x}_i\bm{x}_i^{\top}\right\|_2 =
\left\|\sum_{i\in[p]}\hat{z}_i\left(1-\hat{z}_i\right)\bm{x}_i\bm{x}_i^{\top}\right\|_2 \\
&\leq \left\|\sum_{i\in[p]}\hat{z}_i\bm{x}_i\bm{x}_i^{\top}\right\|_2
\leq \theta_k,
\end{align*}
where 
the first inequality is due to triangle inequality and $\|\bm{x}_i\|_2^2=1$ for each $i\in [p]$, the second inequality is due to $1-\hat{z}_i\in [0,1]$ for all $i\in [p]$ and the last one is due to
\begin{align*}
&\max_{\bm{z}\in [0,1]^p}\left\{\sigma_{\max}\left(z_i\bm{x}_i\bm{x}_i^{\top}\right):\sum_{i\in[p]}z_i=k\right\}\\
&=\max_{\bm{z}\in \{0,1\}^p}\left\{\sigma_{\max}\left(z_i\bm{x}_i\bm{x}_i^{\top}\right):\sum_{i\in[p]}z_i=k\right\}:=\theta_k.
\end{align*}

Now by Lemma~\ref{lemm_trop} with $\sigma_{\min}(\bm{\Sigma}_*)$ denoting the smallest eigenvalue of $\bm{\Sigma}_*$ and $t=\epsilon \sigma_{\min}(\bm{\Sigma}_*)$, we have
\begin{align*}
\Pr\left\{\left\|\sum_{i\in [p]}\left(\hat{\bm\Sigma}-\bm{\Sigma}_*\right)\right\|_2\geq \epsilon \sigma_{\min}(\bm{\Sigma}_*)\right\}\leq n\exp\left(-\frac{\epsilon^2 \sigma_{\min}^2(\bm{\Sigma}_*)}{2\theta_k+2/3\epsilon \sqrt{\theta_1} \sigma_{\min}(\bm{\Sigma}_*)}\right).
\end{align*}
We would like to ensure that the right-hand side of above inequality is at most $\frac{\alpha}{2}$.
Thus,
\begin{align*}
&\Pr\left\{\left\|\sum_{i\in [p]}\left(\hat{\bm\Sigma}-\bm{\Sigma}_*\right)\right\|_2\geq \epsilon \sigma_{\min}(\bm{\Sigma}_*)\right\}\leq \frac{\alpha}{2},\\
(\Leftarrow)\quad&n\exp\left(-\frac{\epsilon^2 \sigma_{\min}^2(\bm{\Sigma}_*)}{2\theta_k+2/3\epsilon \sigma_{\min}(\bm{\Sigma}_*)}\right) \leq \frac{\alpha}{2},\\
(\Leftarrow)\quad &\sigma_{\min}(\bm{\Sigma}_*)\geq \frac{\log(2n/\alpha)\sqrt{\theta_1}}{3\epsilon}+\frac{\sqrt{2\theta_k\log(2n/\alpha)}}{2\epsilon},\\
(\Leftarrow)\quad &\lambda\geq \frac{\log(2n/\alpha)\sqrt{\theta_1}}{3n\epsilon}+\frac{\sqrt{2\theta_k\log(2n/\alpha)}}{2n\epsilon},
\end{align*}
where the second implication is because the following quadratic inequality $a t^2-bt-c\geq 0$ with $a,b,c>0$ is satisfied if $t\geq \frac{b}{a}+\sqrt{\frac{c}{a}}$, and the third implication is due to
$\lambda n\leq \sigma_{\min}(\bm{\Sigma}_*)$.

Then the conclusion follows directly by Weyl's theorem \citep{franklin2012matrix,weyl1912asymptotische}.
\qed}

Based on Lemma \ref{lem1}, we can imply the following bi-criteria approximation of \eqref{eq_model}.
\begin{Theorem}Let $(S,\tilde{\bm z})$ be the output of Algorithm~\ref{alg_rand_round}. Given that $\alpha\in (0,1)$ and \[\lambda\geq \frac{\log(2n/\alpha)\sqrt{\theta_1}}{3n\epsilon}+\frac{\sqrt{2\theta_k\log(2n/\alpha)}}{2n\epsilon},\]
then with probability at least $1-\alpha$, we have
\begin{align}
v^R:=\lambda \bm{y}^{\top}\left[\lambda n\bm{I}_{n}+ \sum_{i\in [p]}\tilde z_i\bm{x}_i\bm{x}_i^{\top}\right]^{-1} \bm y \leq (1+\epsilon)v^*\label{eq_opt_bnd}
\end{align}
and
\begin{align}
\sum_{i\in [p]}\tilde z_i \leq \left(1+\sqrt{\frac{3\log(2/\alpha)}{k}}\right)k.\label{eq_card}
\end{align}
\end{Theorem}
\proof Note that \eqref{eq_opt_bnd} follows from Lemma \ref{lem1}. The result in \eqref{eq_card} holds due to the Chernoff bound \citep{chernoff1952measure}, i.e.,
\begin{align*}
\Pr\left\{\sum_{i\in [p]}\tilde z_i \leq \left(1+\sqrt{\frac{3\log(2/\alpha)}{k}}\right)k\right\} \geq 1-e^{-\frac{\left(\sqrt{\frac{3\log(2/\alpha)}{k}}\right)^2k}{3}}\geq 1-\frac{\alpha}{2}.
\end{align*}
Therefore, by Boole's inequality, we arrive at the conclusion.
\qed

{When revising this paper, we realized a very interesting paper \cite{pilanci2015sparse}, which also studied the same randomized rounding algorithms. Our results distinguish from the work in \cite{pilanci2015sparse} through two aspects: (i) We propose a second order conic program to obtain the continuous relaxation solution, while \cite{pilanci2015sparse} proposed a gradient decent method to solve it; and (ii) Our approximation ratio is multiplicative and does not depend on $p$, while theorem 3 in \cite{pilanci2015sparse} derived an additive approximation bound, which is proportional to the square root of support of the continuous relaxation solution and thus can be $O(\sqrt{p})$. That is, using our notation, our approximation ratio is
\[v^{R} \leq \left(1+ \frac{\log(2n/\alpha)\sqrt{\theta_1}}{3n\lambda}+\frac{\sqrt{2\theta_k\log(2n/\alpha)}}{2n\lambda}\right)v^*\]
and the approximation bound $v^{p}$ in \cite{pilanci2015sparse} is
\[v^{p} -v^*\leq c_4\frac{\sqrt{r\log(\min\{r,n\})}}{n\lambda}\]
where $r=\|\hat{\bm{z}}\|_0$ with $\hat{\bm z}$ denoting the continuous relaxation solution, and $c_4$ is a ``sufficient large constant." Clearly, if $c_4$ is very large or $\|\hat{\bm z}\|_0$ is close to $p$, then our bound is much tighter than \cite{pilanci2015sparse}.
}


Next, let $\bm{\beta}^R$ be the estimator from \Cref{alg_rand_round}, which can be computed according to \eqref{eq_estimator_S} by letting $S$ be the output from \Cref{alg_rand_round}.
Then we can show that the distance between $\bm{\beta}^R$ and $\bm\beta^*$ (i.e., $\|\bm{\beta}^R-\bm\beta^*\|_2$) can be also quite small, where $\bm\beta^*$ is an optimal solution to \eqref{eq_model}.
\begin{Theorem}Let $\bm\beta^*$ be an optimal solution to \eqref{eq_model} with set of selected features $S^*$ and $\bm{\beta}^R$ be the estimator from \Cref{alg_rand_round} with set of selected features $S^R$. Given $\alpha \in (0,1)$, if $\lambda\geq \frac{\log(2n/\alpha)\sqrt{\theta_1}}{3n\epsilon}+\frac{\sqrt{2\theta_k\log(2n/\alpha)}}{2n\epsilon}$, 
then with probability at least $1-\alpha$, we have
\begin{align*}
\|\bm{\beta}^R-\bm\beta^*\|_2 \leq
\frac{\sqrt{ 4 n \theta_{|S^R\setminus S^*|}v^*} } { n\lambda + \sigma_{\min}( \bm{X}_{S^R\cup S^*}^{\top} \bm{X}_{S^R\cup S^*})}
+ \sqrt{\frac{ n \epsilon v^*}{n\lambda+\sigma_{\min}( \bm{X}_{S^R\cup S^*}^{\top} \bm{X}_{S^R\cup S^*})}}.
\end{align*}
\end{Theorem}
\proof
The proof is almost identical to that of \Cref{eq_bound_beta_S}, thus is omitted here.
\qed

Finally, we remark that we can integrate the greedy approach with the randomized algorithm, which is to apply the greedy approach based upon the support of the continuous relaxation solution of \eqref{eq_model2_persp2}.
That is, given that $\hat{\bm z}$ is the optimal solution to \eqref{eq_bound_v2} and $\delta>0$ is a positive constant,
then we first let set $\C:=\left\{i\in [p]:\hat{z}_i\geq \delta\right\}$ and apply greedy approach (\Cref{greedy_approach}) to set $\C$ rather than $[p]$, which could save a significant amount of computational time, in particular when continuous relaxation solution $\hat{\bm z}$ is very sparse.
The detailed description can be found in \Cref{restricted_greedy_approach}.

 \begin{algorithm}[htbp]
 \caption{Proposed Restricted Greedy Approach}
 \label{restricted_greedy_approach}
 \begin{algorithmic}[1]
 \State Let $\hat{\bm z}$ be the optimal solution to \eqref{eq_bound_v2}
 \State Initialize $\delta>0$ (e.g., $\delta=0.01$), $\C:=\left\{i\in [p]:\hat{z}_i\geq \delta\right\}$
 \State Let $S=\emptyset$ and $\bm{A}_S=n\lambda \bm{I}_{n}$
 \For{$i=1,\ldots,k$}
 \State Let $j^*\in \arg\min_{j\in \C\setminus S}\left\{ -\frac{\lambda\left(\bm{y}^{\top}\bm{A}_S^{-1}\bm x_j\right)^2}{1+\bm x_j^{\top}\bm{A}_S^{-1}\bm x_j}\right\}$
 \State Let $S=S\cup\{j^*\}$ and $\bm{A}_S=\bm{A}_S+\bm x_{j^*}\bm x_{j^*}^{\top}, \bm{A}_S^{-1}=\bm{A}_S^{-1}-\frac{\bm{A}_S^{-1}\bm x_{j^*}\bm x_{j^*}^{\top}\bm{A}_S^{-1}}{1+\bm x_{j^*}^{\top}\bm{A}_S^{-1}\bm x_{j^*}}$
 \EndFor
 \State Output $v^{RG} \leftarrow \lambda \bm{y}^{\top}\bm{A}_S^{-1}\bm y$.
 \end{algorithmic}
 \end{algorithm}

{
\section{Selection of Tuning Parameter and Generalization to Sparse Matrix Estimation}\label{sec_generalization}
In this section, we will discuss how to select the tuning parameter $\lambda$ using generalized cross validation and show that our proposed approaches can be extended to sparse matrix estimation.
\subsection{Selection of Tuning Parameter by Generalized Cross Validation (GCV)}\label{sec_gcv}
For a given $k$, we can adopt the commonly-used generalized cross-validation (GCV) \cite{girard1991asymptotic,van2015lecture} to choose the best $\lambda$ in the ridge regression.
Specifically, the GCV can be defined as
\begin{align}
GCV(\lambda) = \frac{1}{n} \sum_{i=1}^{n} \left ( \frac{y_{i} - \hat{y}_{i}}{1 - (\bm{H}_{S})_{ii}} \right )^{2},
\end{align}
where $\bm{H}_{S} = \bm{X}_S (\bm{X}_S^{\top} \bm{X}_S+ n\lambda I)^{-1} \bm{X}_S^{\top}$ denotes the hat matrix of the ridge regression and $\hat{\bm y}=\bm{H}_{S} \bm y$ is the vector of the fitted responses.
With a sequence of $\lambda$ values in $\{\lambda_{1}, \ldots, \lambda_{m}\}$, we can choose the one having the smallest $GCV(\lambda)$ value. {It is worth mentioning that the original GCV \cite{girard1991asymptotic,van2015lecture}  was proposed for the ridge regression without sparsity requirement, and thus GCV used in this paper is a heuristic procedure for the sparse ridge regression problem.}

\subsection{Generalization to Sparse Matrix Estimation}

In this subsection, we consider a sparse matrix estimation proposed by \cite{cai2011constrained}. In that problem, the authors were trying to estimate the inverse of covariance matrix $\hat{\bm\Sigma}\in \Re^{t\times t}$ and choose the sparest estimator. In their model, they optimize the $L_1$ norm of the estimator given that the estimation error is within a constant. Similar to \eqref{eq_model}, instead we can directly optimize the estimation error given that only $k$ sparse elements can be chosen, which can be formulated as below
\begin{align}
v^*=\min_{\bm\Omega}\left\{\|\bm I_{t}-\hat{\bm\Sigma}\Omega\|_F^2+\lambda \|\bm\Omega\|_F^2: \|\bm\Omega\|_0\leq k\right\}, \label{eq_model_matrix}
\end{align}

To view this model as a special case of \eqref{eq_model}, we rewrite matrix $\Omega$ as a vector $\bm\beta\in \Re^{t^2\times 1}$ and $\hat{\bm\Sigma}$ as $\bm X \in \Re^{t\times t^2}$, where
\begin{align*}
&\beta(j+t(i-1))=\Omega(i,j), \forall i, j\in [t], \\
&X(s,r)=\bigg\lbrace\begin{array}{cc}
\Sigma (s, r-t(s-1))& \text{ if } 1\leq r-t(s-1) \leq t\\
0,& \text{ otherwise }
\end{array}, \forall s\in [t], r\in [t^2],\\
& y_{j+t(i-1)}=\begin{cases}
1, &\text{ if }i=j,\\
0,& \text{ otherwise},
\end{cases}\forall i,j\in [t]
\end{align*}
Thus, \eqref{eq_model_matrix} reduces to \eqref{eq_model}. Then the results for sparse ridge regression in the previous sections hold for \eqref{eq_model_matrix}.
}

\section{Experimental Verification}\label{sec_numerical}

{In this section, we illustrate the different algorithms proposed in this paper and how to choose the tuning parameters. Particularly, Section 6.1. focuses on a comparison of  branch and cut algorithm in \cite{bertsimas2017sparse}, MISOC Formulation \eqref{eq_model2_persp2},
heuristic \Cref{bisec-heuristic} in \cite{ahmed2014nonanticipative},
greedy \Cref{greedy_approach}, randomized \Cref{alg_rand_round} and restricted greedy \Cref{restricted_greedy_approach}, Section 6.2. focuses on MISOC Formulation \eqref{eq_model2_persp2} and greedy \Cref{greedy_approach} to illustrate that although fast and close to optimality, greedy \Cref{greedy_approach} might be able to provide near-optimal solutions, and Section 6.3. demonstrates how to choose the tuning parameter $\lambda$ using GCV via a real-world application. The code of greedy algorithm can be found in \url{https://github.com/xwj06/Sparse_Ridge_Regression.git}.}

\subsection{Comparison of  Branch and Cut Algorithm in \cite{bertsimas2017sparse}, MISOC Formulation \eqref{eq_model2_persp2},
Heuristic \Cref{bisec-heuristic} in \cite{ahmed2014nonanticipative},
Greedy \Cref{greedy_approach}, Randomized \Cref{alg_rand_round} and Restricted Greedy \Cref{restricted_greedy_approach} via Large-scale Synthetic Datasets}In this subsection, we conduct experimental studies to evaluate the performance of the proposed methods in comparison with several existing ones on solving sparse ridge regression problems.
The data are generated from the linear model
\begin{align*}
 y = \bm x^{\top} \bm \beta^{0} +\tilde\epsilon,
\end{align*}
where $\tilde\epsilon \sim N(0, \sigma^{2})$. The i.i.d.~samples of $\bm x$ are generated from a multivariate normal distribution with
\begin{align*}
\bm x_{i} \sim N(0, \bm \Sigma), \ i =1, \ldots, n,
\end{align*}
where $\bm \Sigma$ is the covariance matrix with $\sigma_{ij} = \rho^{|i-j|}$ for each $i,j\in [p]$, and $\rho = 0.5$.
{The first $k$ entries of  $\bm \beta^{0} = (\beta_{1}^{0}, \ldots, \beta_{p}^{0})^{\top}$ are nonzero, and their values are drawn randomly from the uniform distribution $\text{Unif}(-3, 3)$}.
To control the signal-to-noise ratio (SNR), we choose the value of $\sigma^{2}$ such that $\text{SNR} = \text{var}(\bm x^{\top} \bm \beta^{0}) / \text{var}(\tilde\epsilon) = 9$.
By generating an i.i.d.~sample of noise $\tilde\epsilon_{1}, \ldots, \tilde\epsilon_{n}$ with $\tilde\epsilon_{i}\sim N(0, \sigma^{2})$ for each $i\in [n]$,
we simulate the response values, i.e., $y_{i} = \bm x_{i}^{\top} \bm \beta^{0} + \tilde\epsilon_{i}$ for each $i\in [n]$.

Recall that the goal is to find a best $k$-sparse estimator for a given $k$.
The performances of the methods in comparison are evaluated by the selection accuracy and computational time.
Here we consider different combinations of $k, n, p$ to generate the simulation data,
where $p \in \{1000, 5000\}$, $n \in \{ 500, 1000, 5000 \}$ and $k \in \{10, 20, 30\}$.
Each simulation setting is repeated by 10 times, i.e., for each tuple $(k,n,p)$, we generate 10 repetitions\footnote{We restrict the simulation to 10 replications because certain existing methods are very slow in computation.}.
For simplicity, for all the testing instances, we set the tuning parameter $\lambda=0.08$. 

The methods in comparison include the branch-and-cut algorithm proposed by \cite{bertsimas2017sparse} based on \eqref{eq_model6}\footnote{Please note that \cite{bertsimas2017sparse} proposed a sophisticated warm-start procedure. However, for the sake of fair comparison, we directly {implemented} branch-and-cut algorithm without any warm-start procedure.},
directly solving \eqref{eq_model2_persp2},
the heuristic \Cref{bisec-heuristic} in \cite{ahmed2014nonanticipative},
the proposed greedy \Cref{greedy_approach}, the proposed randomized \Cref{alg_rand_round} and the proposed restricted greedy \Cref{restricted_greedy_approach}.
Note that the heuristic \Cref{bisec-heuristic} in \cite{ahmed2014nonanticipative} is similar to the LASSO in the use of $L_{1}$ norm to achieve the sparsity.
The commercial solver Gurobi 7.5 with its default setting is used to solve \eqref{eq_model2_persp2} and its continuous relaxation.
We set the time limit to be an hour (3600 seconds).
Due to out-of-memory and out-of-time-limit issues, in the case of $p=5000$, we only compute two of the most effective algorithms: the proposed greedy \Cref{greedy_approach} and the proposed restricted greedy \Cref{restricted_greedy_approach}. 
The comparison results are listed in \Cref{table1} to \Cref{table3}, where the \textit{Avg.~Obj.~Value}, \textit{Avg.~Gap}, \textit{Avg.~Comp.~Time}, and \textit{Avg.~False~Alarm~Rate} denotes
the average objective function value, average optimality gap (of exact methods) from Guorbi, average computational time (in seconds), and average percent of falsely detected features, respectively. For most of the test instances, the optimal value $v^*$ {can be very difficult} to obtain. Therefore, we only compare the objective function values of different algorithms, where the smaller objective function value implies that the output of the algorithm is more accurate.
All the computations were executed on a MacBook Pro with a 2.80 GHz processor and 16GB RAM.

\begin{table}[ht]
\centering
\footnotesize
\setlength\tabcolsep{1pt}
\caption{Comparison of the Branch and Cut algorithm in \cite{bertsimas2017sparse} and directly solving \eqref{eq_model2_persp2} with $p=1000$}
\label{table1}
\begin{tabular}{c|c|c|rrrr|rrrr}
\hline
\multirow{2}{*}{$p$} & \multirow{2}{*}{$k$} & \multirow{2}{*}{$n$} & \multicolumn{4}{c|}{Branch and Cut Algorithm} & \multicolumn{4}{c}{Solving \eqref{eq_model2_persp2}} \\ \cline{4-11}
& & & \begin{tabular}[c]{@{}c@{}}Avg. Obj. \\ Value\end{tabular} & \begin{tabular}[c]{@{}c@{}}Avg. Comp.\\ Time(s) \end{tabular}& \begin{tabular}[c]{@{}c@{}}Avg.\\ Gap \end{tabular} & \begin{tabular}[c]{@{}c@{}}Avg. False \\Alarm Rate\end{tabular} & \begin{tabular}[c]{@{}c@{}}Avg. Obj. \\Value \end{tabular} & \begin{tabular}[c]{@{}c@{}}Avg. Comp.\\ Time(s) \end{tabular}& \begin{tabular}[c]{@{}c@{}}Avg. \\Gap \end{tabular} & \begin{tabular}[c]{@{}c@{}}Avg. False \\Alarm Rate\end{tabular} \\ \hline
\multirow{9}{*}{1000} & \multirow{3}{*}{10} & 500 & 9.71 & 3438.51 & 47.2\% & 26.0\% & 6.83 & 3505.82 & 7.1\% & 5.0\% \\
& & 1000 & 7.11 & 2451.47 & 10.4\% & 5.0\% & 7.27 & 3562.61 & 9.7\% & 7.0\% \\
& & 5000 & NA* & NA & NA & NA & 6.67 & 387.44 & 0.0\% & 0.0\% \\ \cline{2-11}
& \multirow{3}{*}{20} & 500 & 23.02 & 3600.00 & 141.5\% & 45.0\% & 11.98 & 3600.00 & 21.4\% & 20.0\% \\
& & 1000 & 31.52 & 3600.00 & 131.2\% & 50.5\% & 11.55 & 3600.00 & 11.7\% & 18.0\% \\
& & 5000 & NA & NA & NA & NA & 11.30 & 2434.64 & 0.3\% & 0.5\% \\ \cline{2-11}
& \multirow{3}{*}{30} & 500 & 39.62 & 3600.00 & 189.3\% & 51.3\% & 20.42 & 3600.00 & 31.4\% & 27.0\% \\
& & 1000 & 50.63 & 3600.00 & 175.9\% & 55.0\% & 19.16 & 3600.00 & 18.1\% & 22.3\% \\
& & 5000 & NA & NA & NA & NA & 17.79 & 3600.00 & 1.3\% & 5.0\% \\ \cline{1-11}
\end{tabular}\\
\vspace{6pt}
$^{*}$ The NA represents for out of memory instances.
\end{table}

\Cref{table1} reports the comparison results between directly solving \eqref{eq_model2_persp2} and the branch-and-cut algorithm based upon \eqref{eq_model6}.
It is seen that directly solving \eqref{eq_model2_persp2} outperforms the branch-and-cut algorithm for most of the instances, in particular when $k$ becomes large.
This is because (i) we proved in \Cref{thm_bound_comparison} that continuous relaxations of \eqref{eq_model6} and \eqref{eq_model2_persp2} are equivalent,
thus directly solving \eqref{eq_model2_persp2} should perform at least as good as branch and cut algorithm;
and (ii) the branch-and-cut algorithm needs to compute the gradient of the objective function in \eqref{eq_model6}, which involves a very time-consuming $n\times n$ matrix inversion.
However, for both approaches, they reach the time limit for most of the cases, and the average false alarm rates are higher than the approximation algorithms in \Cref{table2}.
Therefore, for large-scale instances, these approaches might not be very desirable.
\begin{table}[h]
\centering
\small
\setlength\tabcolsep{1.0pt}
\caption{Comparison of Heuristic \Cref{bisec-heuristic} in \cite{ahmed2014nonanticipative}, Greedy \Cref{greedy_approach}, Randomized \Cref{alg_rand_round} and Restricted Greedy \Cref{restricted_greedy_approach} with $p=1000$}
\label{table2}
\begin{tabular}{c|c|c|rrr|rrr}
\hline
\multirow{2}{*}{$p$} & \multirow{2}{*}{$k$} & \multirow{2}{*}{$n$} & \multicolumn{3}{c|}{Heuristic \Cref{bisec-heuristic} in \cite{ahmed2014nonanticipative}} & \multicolumn{3}{c}{Proposed Greedy \Cref{greedy_approach}} \\ \cline{4-9}
& & & \begin{tabular}[c]{@{}c@{}}Avg. Obj. \\Value\end{tabular}  &  \begin{tabular}[c]{@{}c@{}}Avg. Comp. \\Time(s)\end{tabular} & \begin{tabular}[c]{@{}c@{}}Avg. False\\ Alarm Rate\end{tabular} &  \begin{tabular}[c]{@{}c@{}}Avg. Obj. \\Value\end{tabular} &  \begin{tabular}[c]{@{}c@{}}Avg. Comp. \\Time(s)\end{tabular}& \begin{tabular}[c]{@{}c@{}}Avg. False \\Alarm Rate\end{tabular} \\ \hline
\multirow{9}{*}{1000} & \multirow{3}{*}{10} & 500 & 9.59 & 579.36 & 3.0\% & 6.60 & 0.47 & 0.0\%\\
& & 1000 & 7.88 & 45.78 & 0.0\% & 6.54 & 0.59 & 0.0\% \\
& & 5000 & 7.24 & 737.06 & 0.0\% & 6.67 & 1.41 & 0.0\% \\ \cline{2-9}
& \multirow{3}{*}{20} & 500 & 15.87 & 589.66 & 14.5\% & 10.86 & 0.79 & 9.0\% \\
& & 1000 & 13.42 & 47.92 & 11.5\% & 10.91 & 2.02 & 4.0\% \\
& & 5000 & 12.66 & 738.55 & 4.5\% & 11.30 & 2.37 & 0.0\% \\ \cline{2-9}
& \multirow{3}{*}{30} & 500 & 28.87 & 583.98 & 17.0\% & 16.88 & 1.13 & 10.7\% \\
& & 1000 & 23.53 & 43.92 & 12.7\% & 17.19 & 1.43 & 6.7\% \\
& & 5000 & 19.74 & 678.10 & 6.0\% & 17.74 & 3.28 & 2.0\% \\ \hline\hline
\multirow{2}{*}{$p$} & \multirow{2}{*}{$k$} & \multirow{2}{*}{$n$} & \multicolumn{3}{c|}{Proposed Randomized \Cref{alg_rand_round}} & \multicolumn{3}{c}{Proposed Restricted Greedy \Cref{restricted_greedy_approach}} \\ \cline{4-9}
& & &  \begin{tabular}[c]{@{}c@{}}Avg. Obj. \\Value\end{tabular} &  \begin{tabular}[c]{@{}c@{}}Avg. Comp. \\Time(s)\end{tabular} & \begin{tabular}[c]{@{}c@{}}Avg. False \\Alarm Rate\end{tabular} &  \begin{tabular}[c]{@{}c@{}}Avg. Obj. \\Value\end{tabular}&  \begin{tabular}[c]{@{}c@{}}Avg. Comp. \\Time(s)\end{tabular} & \begin{tabular}[c]{@{}c@{}}Avg. False \\Alarm Rate\end{tabular} \\ \hline
\multirow{9}{*}{1000} & \multirow{3}{*}{10} & 500 & 7.79 & 4.06 & 14.0\% & 6.60 & 3.84 & 0.0\% \\
& & 1000 & 6.86 & 11.21 & 6.0\% & 6.54 & 10.58 & 0.0\% \\
& & 5000 & 6.67 & 181.77 & 0.0\% & 6.67 & 186.81 & 0.0\% \\ \cline{2-9}
& \multirow{3}{*}{20} & 500 & 12.88 & 4.01 & 23.5\% & 10.86 & 3.80 & 9.0\% \\
& & 1000 & 11.68 & 10.84 & 18.0\% & 10.91 & 13.81 & 4.0\% \\
& & 5000 & 11.40 & 199.31 & 6.5\% & 11.30 & 202.66 & 0.0\% \\ \cline{2-9}
& \multirow{3}{*}{30} & 500 & 20.89 & 4.21 & 26.3\% & 16.89 & 4.06 & 11.0\% \\
& & 1000 & 19.89 & 10.58 & 24.0\% & 17.19 & 11.94 & 6.7\% \\
& & 5000 & 18.11 & 167.95 & 10.0\% & 17.74 & 170.14 & 2.0\% \\ \hline
\end{tabular}
\end{table}

\begin{table}
\centering
\small
\setlength\tabcolsep{1.0pt}
\caption{Comparison of Greedy \Cref{greedy_approach} and Restricted Greedy \Cref{restricted_greedy_approach} with $p=5000$}
\label{table3}
\begin{tabular}{c|c|c|rrr|rrr}
\hline
\multirow{2}{*}{$p$} & \multirow{2}{*}{$k$} & \multirow{2}{*}{$n$} & \multicolumn{3}{c|}{Proposed Greedy \Cref{greedy_approach}} & \multicolumn{3}{c}{Proposed Restricted Greedy \Cref{restricted_greedy_approach}} \\ \cline{4-9}
& & &  \begin{tabular}[c]{@{}c@{}}Avg. Obj. \\Value\end{tabular} &  \begin{tabular}[c]{@{}c@{}}Avg. Comp. \\Time(s)\end{tabular} & \begin{tabular}[c]{@{}c@{}}Avg. False \\Alarm Rate\end{tabular} &  \begin{tabular}[c]{@{}c@{}}Avg. Obj. \\Value\end{tabular} &  \begin{tabular}[c]{@{}c@{}}Avg. Comp. \\Time(s)\end{tabular} & \begin{tabular}[c]{@{}c@{}}Avg. False \\Alarm Rate\end{tabular} \\ \hline
\multirow{9}{*}{5000} & \multirow{3}{*}{10} & 500 & 4.57 & 2.31 & 0.0\% & 4.57 & 15.81 & 0.0\% \\
& & 1000 & 4.59 & 3.13 & 0.0\% & 4.59 & 39.06 & 0.0\% \\
& & 5000 & 4.68 & 9.04 & 0.0\% & 4.68 & 1451.78 & 0.0\% \\ \cline{2-9}
& \multirow{3}{*}{20} & 500 & 12.86 & 4.31 & 8.0\% & 12.86 & 15.69 & 8.0\% \\
& & 1000 & 13.35 & 5.41 & 2.5\% & 13.35 & 38.14 & 2.5\% \\
& & 5000 & 13.27 & 14.58 & 0.0\% & 13.27 & 1426.93 & 0.0\% \\ \cline{2-9}
& \multirow{3}{*}{30} & 500 & 14.02 & 5.98 & 20.7\% & 14.02 & 16.24 & 20.7\% \\
& & 1000 & 14.97 & 8.21 & 12.7\% & 14.97 & 39.41 & 12.7\% \\
& & 5000 & 15.60 & 20.52 & 3.3\% & 15.60 & 1503.48 & 3.3\% \\ \hline
\end{tabular}
\end{table}

From \Cref{table2} and \Cref{table1}, the proposed greedy \Cref{greedy_approach} and restricted greedy \Cref{restricted_greedy_approach} apparently perform best among all comparison methods.
We see that for the instances with $k=10$, the heuristic \Cref{bisec-heuristic}, greedy \Cref{greedy_approach} and restricted greedy \Cref{restricted_greedy_approach} find almost all the features,
while the randomized \Cref{alg_rand_round} performs slightly worse.
When the number of active features, $k$, grows, all the methods in comparison have relatively larger false alarm rates.
Their performance of identifying right features improves as the sample size $n$ increases, i.e., providing more information.
For the heuristic \Cref{bisec-heuristic} in \cite{ahmed2014nonanticipative}, it is less accurate and takes a much longer time. Thus, it might not be a good option for large-scale instances either.
In contrast, we note that the greedy \Cref{greedy_approach} is much {more} accurate.
It runs very fast with the computation time, which is proportional to $n,p,k$.
But the randomized \Cref{alg_rand_round}, which depends on the solution time of solving the continuous relaxation of \eqref{eq_model2_persp2}, is quite insensitive to $k$ in terms of computation time.
Therefore, by integrating these two together, the restricted greedy \Cref{restricted_greedy_approach} can be advantageous for large $k$, providing accurate estimation with fast computation.
For the numerical study with $p=5000$ below, we choose these two most efficient algorithms for comparison.


In \Cref{table3},
we observe that the greedy \Cref{greedy_approach} and the restricted greedy \Cref{restricted_greedy_approach} have exactly the same false alarm rates.
But the greedy \Cref{greedy_approach} is much faster than the restricted greedy \Cref{restricted_greedy_approach}.
This is mainly because it takes a much longer time to solve the continuous relaxation to the optimality and for these instances, $k$ is relatively small.
In particular, for a large-scale datasets (e.g., $n=p=5000$),
the computation time of the restricted greedy \Cref{restricted_greedy_approach} is much longer time than those in the case with $p=1000$.
But, the greedy \Cref{greedy_approach} can still find very high-quality solutions within 30 seconds of computation time.
On the other hand, we note that the accuracy of both approaches grows when the sample size increases.
Thus, we would recommend finding a reasonable sample size that the greedy methods can work efficiently and identify the features accurately.

{\begin{table}[htbp]
\caption{A comparison with the forward selection algorithm proposed in \cite{hastie2017extended}. Note that the solver in \cite{hastie2017extended} only works for sparse regression. Thus, we only compare the computational time.}
\label{tab:my-table}
\centering\small
\begin{tabular}{|l|l|l|r|r|}
\hline
$p$ & $k$ & $n$ & \begin{tabular}[c]{@{}r@{}}Greedy \Cref{greedy_approach} \\ Time (s)\end{tabular} & \begin{tabular}[c]{@{}r@{}}Forward Selection in \cite{hastie2017extended}\\ Time (s)\end{tabular} \\ \hline
\multirow{9}{*}{1000} & \multirow{3}{*}{10} & 500 & 0.47 & 0.79 \\ \cline{3-5}
 &  & 1000 & 0.59 & 1.10 \\ \cline{3-5}
 &  & 5000 & 1.41 & 11.41 \\ \cline{2-5}
 & \multirow{3}{*}{20} & 500 & 0.79 & 1.33 \\ \cline{3-5}
 &  & 1000 & 1.01 & 2.86 \\ \cline{3-5}
 &  & 5000 & 2.37 & 26.49 \\ \cline{2-5}
 & \multirow{3}{*}{30} & 500 & 1.13 & 2.17 \\ \cline{3-5}
 &  & 1000 & 1.43 & 4.01 \\ \cline{3-5}
 &  & 5000 & 3.28 & 38.98 \\ \hline
\end{tabular}
\end{table}

We have numerically compared our implementation with the state-of-art R package posted by \cite{hastie2017extended}. 
Table~\ref{tab:my-table} summarize the comparison in terms of computational time.
It is seen that our implementation can outperform the one in \cite{hastie2017extended}.
The advantage appears to be more striking as $n$ becomes larger.
{Thus, we envision that our implementation for the greedy approach (or forward selection) is efficient and can be interesting to the readers.} }

{
\subsection{Further Investigation of MISOC Formulation \eqref{eq_model2_persp2} and\\ Greedy \Cref{greedy_approach} with Varying SNR and Tuning Parameter $\lambda$ via\\ Medium-size Synthetic Datasets}

Following the same data generating procedure in the previous subsection, we conduct a thorough comparison of MISOC Formulation \eqref{eq_model2_persp2} and Greedy \Cref{greedy_approach}. In particular, we generate 16 instances with $n=100,p=40,k\in \{5,10,15,20\}, \textrm{SNR}\in \{0.5,1,2,4\}$ and to illustrate the effects of tuning parameter $\lambda$, we let it vary from the range $\{0.01,0.1,1,10\}$. Similarly, each simulation setting is repeated 10 times, and the average results are reported in \Cref{table4} and \Cref{table5}.

In \Cref{table4} and \Cref{table5}, it is seen that for these instances, formulation \eqref{eq_model2_persp2} can be solved to optimality within 2 minutes, while greedy \Cref{greedy_approach} can find the very near-optimal solutions within 0.1 second. We also see that in terms of average objective value and average false alarm rate, greedy \Cref{greedy_approach} in this case is slightly worse than formulation  \eqref{eq_model2_persp2}, since the latter is able to provide exact solutions. Thus, if the instances are not large, we suggest solving exact formulation \eqref{eq_model2_persp2}, which indeed provides the best performance. 
As for the SNR, we see that the false alarm rates of both approaches decrease as SNR increases, which is consistent with the intuition since higher SNR implies stronger signal, and thus more accurate prediction. In terms of tuning parameter, we see that the computational time of formulation \eqref{eq_model2_persp2} changes significantly as $\lambda$ increases. On the other hand, if the tuning parameter $\lambda$ is too big, the false alarm rate will increase significantly. Thus, a proper choice of the tuning parameter $\lambda$ will be critical for formulation \eqref{eq_model2_persp2}. In next subsection, we will use the generalized cross validation to choose a proper tuning parameter $\lambda$ for a real-world case.

\begin{table}

\centering

\small

\setlength\tabcolsep{1.0pt}

\caption{Comparison of MISOC Formulation \eqref{eq_model2_persp2}  and Greedy \Cref{greedy_approach} with $n=100,p=40, k\in \{5,10\}$}

\label{table4}

\begin{tabular}{c|c|c|rrr|rrr}

\hline

\multirow{2}{*}{$k$} & \multirow{2}{*}{SNR} & \multirow{2}{*}{$\lambda$} & \multicolumn{3}{c|}{MISOC Formulation \eqref{eq_model2_persp2}} & \multicolumn{3}{c}{Proposed Greedy \Cref{greedy_approach}} \\ \cline{4-9}

& & &  \begin{tabular}[c]{@{}c@{}}Avg. Obj. \\Value\end{tabular} &  \begin{tabular}[c]{@{}c@{}}Avg. Comp. \\Time(s)\end{tabular} & \begin{tabular}[c]{@{}c@{}}Avg. False \\Alarm Rate\end{tabular} &  \begin{tabular}[c]{@{}c@{}}Avg. Obj. \\Value\end{tabular} &  \begin{tabular}[c]{@{}c@{}}Avg. Comp. \\Time(s)\end{tabular} & \begin{tabular}[c]{@{}c@{}}Avg. False \\Alarm Rate\end{tabular} \\ \hline

\multirow{16}{*}{5} & \multirow{4}{*}{0.5} & 0.01 & 29.75 & 81.72 & 44.0\% & 29.88 & 0.013 & 46.0\% \\
 &  & 0.1 & 31.65 & 4.44 & 38.0\% & 31.72 & 0.014 & 38.0\% \\
 &  & 1 & 39.98 & 0.28 & 36.0\% & 39.99 & 0.013 & 36.0\% \\
 &  & 10 & 49.13 & 0.29 & 44.0\% & 49.13 & 0.010 & 44.0\% \\ \cline{2-9}
 & \multirow{4}{*}{1} & 0.01 & 15.07 & 54.24 & 38.0\% & 15.10 & 0.011 & 38.0\% \\
 &  & 0.1 & 16.56 & 2.11 & 38.0\% & 16.57 & 0.012 & 38.0\% \\
 &  & 1 & 23.90 & 0.29 & 44.0\% & 23.90 & 0.014 & 44.0\% \\
 &  & 10 & 32.73 & 0.28 & 52.0\% & 32.73 & 0.014 & 52.0\% \\ \cline{2-9}
 & \multirow{4}{*}{2} & 0.01 & 7.11 & 64.68 & 26.0\% & 7.11 & 0.014 & 26.0\% \\
 &  & 0.1 & 8.40 & 2.80 & 26.0\% & 8.40 & 0.014 & 26.0\% \\
 &  & 1 & 14.71 & 0.29 & 32.0\% & 14.71 & 0.011 & 32.0\% \\
 &  & 10 & 21.90 & 0.28 & 44.0\% & 21.90 & 0.010 & 44.0\% \\ \cline{2-9}
 & \multirow{4}{*}{4} & 0.01 & 3.86 & 20.76 & 20.0\% & 3.86 & 0.010 & 20.0\% \\
 &  & 0.1 & 5.13 & 1.17 & 18.0\% & 5.14 & 0.013 & 20.0\% \\
 &  & 1 & 11.17 & 0.28 & 34.0\% & 11.17 & 0.014 & 34.0\% \\
 &  & 10 & 18.15 & 0.28 & 40.0\% & 18.15 & 0.014 & 40.0\% \\ \cline{2-9}
\multirow{16}{*}{10} & \multirow{4}{*}{0.5} & 0.01 & 49.00 & 43.43 & 49.0\% & 49.58 & 0.025 & 47.0\% \\
 &  & 0.1 & 53.08 & 3.00 & 48.0\% & 53.30 & 0.021 & 46.0\% \\
 &  & 1 & 68.60 & 0.25 & 50.0\% & 68.61 & 0.019 & 51.0\% \\
 &  & 10 & 85.37 & 0.24 & 55.0\% & 85.37 & 0.019 & 55.0\% \\ \cline{2-9}
 & \multirow{4}{*}{1} & 0.01 & 24.62 & 12.44 & 46.0\% & 24.75 & 0.023 & 40.0\% \\
 &  & 0.1 & 27.81 & 0.68 & 40.0\% & 27.89 & 0.024 & 39.0\% \\
 &  & 1 & 41.93 & 0.23 & 44.0\% & 41.94 & 0.025 & 44.0\% \\
 &  & 10 & 58.06 & 0.27 & 50.0\% & 58.06 & 0.022 & 51.0\% \\ \cline{2-9}
 & \multirow{4}{*}{2} & 0.01 & 13.02 & 7.70 & 29.0\% & 13.08 & 0.022 & 29.0\% \\
 &  & 0.1 & 15.84 & 0.51 & 28.0\% & 15.91 & 0.020 & 30.0\% \\
 &  & 1 & 28.53 & 0.21 & 34.0\% & 28.53 & 0.025 & 34.0\% \\
 &  & 10 & 42.95 & 0.25 & 42.0\% & 42.95 & 0.024 & 42.0\% \\ \cline{2-9}
 & \multirow{4}{*}{4} & 0.01 & 6.70 & 2.51 & 27.0\% & 6.75 & 0.024 & 30.0\% \\
 &  & 0.1 & 9.22 & 0.40 & 30.0\% & 9.23 & 0.020 & 30.0\% \\
 &  & 1 & 20.67 & 0.21 & 34.0\% & 20.67 & 0.020 & 34.0\% \\
 &  & 10 & 34.09 & 0.25 & 49.0\% & 34.09 & 0.021 & 49.0\% \\\hline
\end{tabular}
\end{table}

\begin{table}

\centering

\small

\setlength\tabcolsep{1.0pt}

\caption{Comparison of MISOC Formulation \eqref{eq_model2_persp2}  and Greedy \Cref{greedy_approach} with $n=100,p=40, k\in \{15,20\}$}

\label{table5}

\begin{tabular}{c|c|c|rrr|rrr}

\hline

\multirow{2}{*}{$k$} & \multirow{2}{*}{SNR} & \multirow{2}{*}{$\lambda$} & \multicolumn{3}{c|}{MISOC Formulation \eqref{eq_model2_persp2}} & \multicolumn{3}{c}{Proposed Greedy \Cref{greedy_approach}} \\ \cline{4-9}

& & &  \begin{tabular}[c]{@{}c@{}}Avg. Obj. \\Value\end{tabular} &  \begin{tabular}[c]{@{}c@{}}Avg. Comp. \\Time(s)\end{tabular} & \begin{tabular}[c]{@{}c@{}}Avg. False \\Alarm Rate\end{tabular} &  \begin{tabular}[c]{@{}c@{}}Avg. Obj. \\Value\end{tabular} &  \begin{tabular}[c]{@{}c@{}}Avg. Comp. \\Time(s)\end{tabular} & \begin{tabular}[c]{@{}c@{}}Avg. False \\Alarm Rate\end{tabular} \\ \hline

\multirow{16}{*}{15} & \multirow{4}{*}{0.5} & 0.01 & 66.81 & 110.75 & 46.7\% & 68.61 & 0.035 & 48.7\% \\
 &  & 0.1 & 74.92 & 4.54 & 46.7\% & 75.61 & 0.035 & 46.0\% \\
 &  & 1 & 101.07 & 0.27 & 42.0\% & 101.09 & 0.029 & 42.0\% \\
 &  & 10 & 126.31 & 0.27 & 46.7\% & 126.31 & 0.028 & 46.7\% \\ \cline{2-9}
 & \multirow{4}{*}{1} & 0.01 & 34.42 & 162.55 & 37.3\% & 35.15 & 0.033 & 38.0\% \\
 &  & 0.1 & 39.85 & 5.54 & 36.7\% & 40.13 & 0.035 & 36.7\% \\
 &  & 1 & 60.56 & 0.33 & 41.3\% & 60.64 & 0.035 & 42.0\% \\
 &  & 10 & 83.68 & 0.29 & 49.3\% & 83.68 & 0.032 & 49.3\% \\ \cline{2-9}
 & \multirow{4}{*}{2} & 0.01 & 18.85 & 34.03 & 25.3\% & 19.09 & 0.025 & 27.3\% \\
 &  & 0.1 & 23.72 & 1.66 & 25.3\% & 23.88 & 0.028 & 26.0\% \\
 &  & 1 & 42.64 & 0.27 & 32.7\% & 42.69 & 0.036 & 34.0\% \\
 &  & 10 & 61.73 & 0.25 & 41.3\% & 61.73 & 0.040 & 40.7\% \\ \cline{2-9}
 & \multirow{4}{*}{4} & 0.01 & 9.50 & 11.84 & 22.7\% & 9.61 & 0.026 & 23.3\% \\
 &  & 0.1 & 14.01 & 0.53 & 24.0\% & 14.10 & 0.028 & 22.7\% \\
 &  & 1 & 32.67 & 0.26 & 30.0\% & 32.67 & 0.033 & 30.7\% \\
 &  & 10 & 52.66 & 0.28 & 39.3\% & 52.66 & 0.037 & 39.3\% \\ \cline{2-9}
\multirow{16}{*}{20} & \multirow{4}{*}{0.5} & 0.01 & 66.04 & 81.72 & 43.0\% & 66.60 & 0.040 & 43.5\% \\
 &  & 0.1 & 74.61 & 4.44 & 41.5\% & 74.72 & 0.037 & 41.0\% \\
 &  & 1 & 103.88 & 0.28 & 38.0\% & 103.90 & 0.048 & 38.5\% \\
 &  & 10 & 136.87 & 0.29 & 41.5\% & 136.87 & 0.044 & 41.5\% \\ \cline{2-9}
 & \multirow{4}{*}{1} & 0.01 & 28.59 & 54.24 & 34.5\% & 28.88 & 0.035 & 34.5\% \\
 &  & 0.1 & 33.63 & 2.11 & 33.5\% & 33.80 & 0.037 & 32.5\% \\
 &  & 1 & 52.82 & 0.29 & 34.0\% & 52.84 & 0.045 & 35.5\% \\
 &  & 10 & 76.04 & 0.28 & 38.5\% & 76.04 & 0.046 & 38.5\% \\ \cline{2-9}
 & \multirow{4}{*}{2} & 0.01 & 15.95 & 64.68 & 29.0\% & 16.22 & 0.034 & 27.5\% \\
 &  & 0.1 & 20.16 & 2.80 & 27.5\% & 20.27 & 0.039 & 28.0\% \\
 &  & 1 & 36.68 & 0.29 & 27.0\% & 36.68 & 0.051 & 27.5\% \\
 &  & 10 & 58.11 & 0.28 & 35.0\% & 58.11 & 0.052 & 35.5\% \\ \cline{2-9}
 & \multirow{4}{*}{4} & 0.01 & 8.24 & 20.76 & 27.0\% & 8.32 & 0.058 & 24.5\% \\
 &  & 0.1 & 11.78 & 1.17 & 24.0\% & 11.79 & 0.052 & 23.5\% \\
 &  & 1 & 27.25 & 0.28 & 32.0\% & 27.25 & 0.055 & 32.0\% \\
 &  & 10 & 48.07 & 0.28 & 38.0\% & 48.07 & 0.061 & 38.0\%\\\hline
\end{tabular}
\end{table}

\subsection{A Real-world Case Study using the Dataset in \cite{Weber773}}
In this subsection, we conduct a case study using the dataset in \cite{Weber773}, which attempted to map the loci on the third chromosome of Drosophila melanogaster that will influence an index of wing shape. The dataset has $n=701$ recombinant inbred lines (i.e., observations) and genotypes of 48 markers, where 11 markers are highly correlated with others, and are thus removed. The selected 37 markers and their corresponding indices can be found at \url{https://www4.stat.ncsu.edu/~boos/var.select/wing.shape.html}. Similar to  \cite{Weber773}, we also consider the interactions of the remaining 37 markers and thus, there are $p=37+\binom{37}{2}=703$ features in total. We use generalized cross validation to choose a proper tuning parameter $\lambda$ from the list $\{10^{-5}, 10^{-4},10^{-3},0.01,0.1,0.2,0.5,1\}$ for each $k\in\{10,20,40\}$. We use greedy \Cref{greedy_approach} to solve all the instances and the total running time is within 1 minutes. \Cref{table6} shows the feature selection results.

In \Cref{table6} , we see that using GCV procedure in Section \ref{sec_gcv}, the best tuning parameter $\lambda$ tends to be small in particular when $k$ increases. In general, a proper $k$ can be determined by biologists or engineers, and as long as $k$ is not very large, we are able to deliver near-optimal feature selections efficiently. In fact, we see that if $k=40$, then we can identify all the necessary markers listed in the table 3 of \cite{mak2019cmenet} except x23. This demonstrates that our proposed method is indeed effective for feature selection problems.
\begin{table}[]
\caption{Feature Selection Results using the Dataset in \cite{Weber773} and GCV in \Cref{sec_gcv}. Here, x$i$ denotes $i$th marker and x$i$.x$j$ represents the interaction of markers $i$ and $j$.}
\centering
\small
\setlength\tabcolsep{1.0pt}
\label{table6}
\begin{tabular}{|l|l|l|}
\hline
$\lambda$ & $k$ & Selected Features \\ \hline
$10^{-4}$ & 10 & x1, x18, x48, x1.x18, x1.x48, x5.x15, x11.x42, x16.x33, x17.x48, x42.x45 \\ \hline
$10^{-5}$ & 20 & \begin{tabular}[c]{@{}l@{}}x1, x18, x37, x48, x1.x4, x1.x18, x1.x48, x5.x15, x11.x42, x14.x37, \\ x16.x33, x16.x45, x17.x27, x17.x48, x34.x40, x34.x48, x36.x40, x36.x48, \\ x40.x45, x42.x45\end{tabular} \\ \hline
$10^{-5}$ & 40 & \begin{tabular}[c]{@{}l@{}}x1, x10, x18, x37, x40, x48, x1.x4, x1.x10, x1.x18, x1.x48, x3.x44, x5.x15, \\ x5.x48, x7.x10, x9.x10, x9.x13, x9.x18, x10.x13, x10.x18, x10.x30, x11.x40, \\ x11.x42, x12.x36, x13.x33, x14.x37, x16.x33, x16.x45, x17.x27, x17.x48, \\ x18.x36, x34.x40, x34.x45, x34.x48, x35.x45, x35.x48, x36.x40, x36.x48, \\ x40.x45, x42.x45, x46.x48\end{tabular} \\ \hline
\end{tabular}
\end{table}

}

\section{Conclusion}\label{sec_conclusion}
This paper studies the sparse ridge regression with the use of exact $L_{0}$ norm for the sparsity.
It is known that imposing $L_{0}$ norm for the sparsity in regression can often become an NP-hard problem in variable selection and estimation.
We {present} a mixed integer second order conic (MISOC) formulation, which is big-M free and is based on perspective formulation.
We prove that the continuous relaxation of this MISOC reformulation is equivalent to the convex integer program (CIP) formulation studied by literature, and can be stronger than straightforward big-M formulation.
Based on these two formulations, we propose two scalable algorithms, the greedy and randomized algorithms, for solving the sparse ridge regression.
Under mild conditions, both algorithms can find near-optimal solutions with performance grantees. Our numerical study demonstrates that the proposed algorithms can indeed solve large-scale instances efficiently. {In general, we recommend solving MISOC formulation first, which might be efficient; otherwise, using the scalable algorithms {studied} in this paper, which has the performance guarantees.}


{\section*{Acknowledgment} We appreciate two anonymous referees and the associate editor for their valuable comments for improving this paper.}

\bibliography{Reference}
\end{document}